\documentclass[times,onecolumn,nopreprintline]{elsarticle}
\usepackage[left=2cm,right=2cm,top=2cm,bottom=2cm]{geometry}
\usepackage[english]{babel}
\usepackage{graphicx}
\usepackage{epsfig}
\usepackage{ae}
\usepackage{subfigure}
\usepackage{url}
\usepackage{svn}
\usepackage[nolist]{acronym}
\usepackage{amsfonts}
\usepackage{amsmath}
\usepackage{amssymb}
\usepackage{amstext}
\usepackage{amsthm}
\usepackage{xspace}
\usepackage{pstricks}
\usepackage{pst-node}
\usepackage{pst-plot}
\usepackage{multirow}
\usepackage[utf8]{inputenc}
\graphicspath{{figure/}} 
\usepackage{multirow}
\usepackage{eurosym}
\usepackage{lscape}
\biboptions{sort&compress}
\usepackage{picins}

\usepackage{setspace}
\begin{document}
\begin{frontmatter}
\title{Energy management in communication networks: a journey through modelling and optimization glasses}
\author[turin]{B. Addis}
\ead{bernardetta.addis@loria.fr}
\author[milan]{A. Capone \corref{cor1}}
\ead{capone@elet.polimi.it}
\author[milan]{G. Carello}
\ead{giuliana.carello@polimi.it}
\author[milan,montreal]{L.G. Gianoli}
\ead{luca-giovanni.gianoli@polymtl.ca}
\author[montreal]{B. Sans\`o}
\ead{brunilde.sanso@polymtl.ca}
\cortext[cor1]{Corresponding author}
\address[milan]{Politecnico di Milano, Dipartimento di Elettronica, Informazione e Bioingegneria, Italy}
\address[turin]{LORIA (UMR 7503 CNRS), Universit\'e de Lorraine, INRIA Nancy-Grand Est, France}
\address[montreal]{\'Ecole Polytechnique de Montr\'eal, D\'epartement de G\'enie \'Electrique, Canada}
\date{}

\begin{abstract}
The widespread proliferation of Internet and wireless applications has produced a significant increase of ICT energy footprint. As a response, in the last five years, significant efforts have been undertaken to include energy-awareness into network management. Several green networking frameworks have been proposed  by carefully managing the network routing and the power state of network devices. 

Even though approaches proposed differ based on network technologies and sleep modes of nodes and interfaces, they all aim at tailoring the active network resources to the varying traffic needs in order to minimize energy consumption. From a modeling point of view, this has several commonalities with classical network design and routing problems, even if with different objectives and in a dynamic context.

With most researchers focused on addressing the complex and crucial technological aspects of green networking schemes, there has been so far little attention on understanding the modeling similarities and differences of proposed solutions. This paper fills the gap surveying the literature with optimization modeling glasses, following a tutorial approach that guides through the different components of the models with a unified symbolism. A detailed classification of the previous work based on the modeling issues included is also proposed.

\vspace{0.3cm}

\end{abstract}


\end{frontmatter}

\section{Introduction}
The widespread of ICT has caused a significant increase of its energy footprint. In fact,
the ICT sector alone produces at least 2\% (0.8 Gt CO$_{2}$ per year) of global GHG emissions~\cite{toure08,theclimategroup08,forster09}, exceeding even the amount of the aviation sector~\cite{ajmonemarsan09}. Some studies \cite{vereecken08} expect this amount to reach the impressive value of 1.4 Gt of CO$_{2}$ per year by the end of 2020, which would account for approximately 2.8\% of global emissions. 
According to 2007 figures, ICT power requirement was estimated to be within a range from 2\% to 10\% of global power consumption~\cite{lubritto08,koomey07}, while the energy demand of network equipment, excluding servers in data centers, was around 22 GW with expectations of reaching 95 GW in 2020 \cite{vereecken08}. % while the overall yearly consumption of broadband modems and routers was 35 TWh/y \citep{malmodin10}. 
Other data related to energy consumption shows that telecom operators demand grew from 150 Twh/y in 2007 to 260 TWh/y in 2012, an amount that accounts for 3\% of the total worldwide need \cite{lambert12}. Finally, other studies focusing on single Internet Service Provider (ISP) show that the energy consumed by the largest providers such as AT\&T or China Mobile reached 11 TWh per year in 2010, while in the case of medium sized operators like Telecom Italia and GRNET the figures are expected to approach 400 GWh in 2015 \cite{bolla12}.

Although green techniques have been applied to almost any aspect of ICT with the aim of improving its energy efficiency, this survey specifically focuses on energy-aware management of IP networks. It is worth pointing out that for ISPs,  about 20\% of power consumption is due to the core/backbone, while the remaining 75\% is consumed by the access/edge ones \cite{bolla10,bolla11}. However, technology upgrades and increasing transmission rates are expected to increase the backbone portion up to 40\% in 2017 \cite{lange09} and 50\% in 2020 \cite{lange11,hinton11}. 

%Note that the survey deliberately excludes the WDM infrastructure because, dues to the large body of literature covering this specific technology,
%we think that a separate survey should deal with it.

%(i) the IP/MPLS layer is responsible for at least 60\% of the whole consumption of a wired network, while the optical layer is by far more energy efficient, with a energy demand quota around 10\% of that of the whole network \cite{vereecken11}; (ii) from the modeling perspective, energy-aware network management operated at the optical layer is substantially different from pure IP network management and, due to the large body of work dealing with it, we believe that another survey should be dedicated to cover this domain.  

In recent years, a very large body of work has been published
in the area of green networking, including a few survey papers. Even though there is a vast and specific literature on IP energy-aware management,
to our knowledge, there has not been so far a thorough review and analysis of all the different modeling features of the related optimization problems.
%Some survey articles on general green networking practices have been recently published. However, although  large body of work on energy-aware network management for wired IP networks, the literature in this specific field has been never thoroughly discussed. 

In this paper we present a tutorial-survey of this hot area of research. We first introduce the reader to the family of \textit{energy-aware network management} (EANM) problems, by discussing step by step all their important modeling features. We then classify previous work based on the model characteristics identified and comment on the open issues.

The paper is organized as follows. We provide a brief technological background and describe the features characterizing different problems within the general EANM framework 
in Section \ref {sec:approaches}. In Section \ref {sec:problems} we then discuss EANM from an optimization modeling perspective, guiding the reader through a series of mathematical programming models representing the different problems addressed in the literature and highlighting the strong ties between energy management problems and the classical network design problems widely studied by the operations research community.
Section \ref {sec:review} is the survey part of the paper, where the state of the art literature on EANM is classified according to different criteria representing some key characteristics of the models used. 
The last Section presents a wrap-up of the work that was carried out in the paper and provides some conclusive remarks.

%%%%%%%%%%%%%%%%%%%%%%%%%%%%%%%%%%

\section{Energy-aware network management: a general overview}\label{sec:approaches}
IP networks can be made greener by working at different levels \cite{mellah09,minami08}: by developing novel energy-efficient devices and architectures which offer support for sleeping primitives  \cite{bolla10a} or \textit{pipeline IP forwarding} \cite{baldi09}, by defining local procedures to autonomously adjust the state of a single network device according to real-time measurements \cite{nedevschi09}, and by coordinating the management of the whole network infrastructure to optimize both energy consumption and performance of both routing and device configurations (EANM) \cite{amaldi13,zhang10,chiaraviglio12a}.

Practically speaking, the goal of EANM is to adapt the network consumption to the  traffic levels. The way EANM is performed relates to the following elements: (i) the power profile of network devices, (ii) the routing protocol, according to which traffic engineering is performed, (iii) the QoS requirements requested by each traffic flow, (iv) when and how frequently network re-configurations should be performed, i.e., in advance off-line or in real-time on-line, (v) where to locate the intelligence of the system (centralized or distributed architecture) and finally, (vi) network survivability in case of failures.

\subsection{Power profiles}
The power profile of a network device is its power consumption versus traffic load. Depending on the shape of the profile, the network can be operated in an energy-efficient manner according to different policies. 

Assuming non-linear power profiles, different approaches can be considered \cite{garroppo13a}. For instance, in case of cubic power profiles, it would be more energy efficient to distribute the traffic among multiple devices to keep the average utilization as low as possible, while in presence of logarithmic profiles it would be much more effective to consolidate the traffic over a very restricted set of network elements.

According to several recent studies, however, current network devices and their main components, e.g. router chassis and line cards, present an almost linear power profile with a pretty high consumption at zero load and relatively small increase up to maximum consumption at full load  \cite{chabarek08,mahadevan09,mellah09,vanheddeghem12b}. 
The power needed to maintain a device active even with no load makes up around 90\% of the energy consumed during peak utilization periods. We refer to the \emph{Powerlib} database \cite{vanheddeghem12d} for an accurate collection of consumption figures for several networking devices, both IP and optical, produced by different vendors. Since in practice the profile is almost flat and energy savings can be achieved only switching off the device (sleep mode), power profiles are often approximated with an ON-OFF (step) curve.

Although great efforts have been made to improve the load proportionality of the energy consumption of next generation network devices (see e.g. \cite{digregorio13}), the ON-OFF profile characterizes the current devices and makes sleeping  the most promising and effective strategy to adapt the consumption of IP networks to the incoming traffic levels. Analytic studies such as \cite{bolla11} and \cite{bolla12} estimate that around 50\% energy savings could be achieved. Furthermore, \cite{chiaraviglio11,chiaraviglio13a,charalampou13} remark that sleeping-based strategies remain effective when applied to network devices characterized by a utilization-proportional power profile provided that fixed and proportional consumption components are of the same order of magnitude.

In addition to pure sleeping strategies, adaptive link rate (ALR) has been recently standardized by the IEEE 802.3az engineering task force~\cite{christensen10} to efficiently adapt the transmission peak rate, and thus the corresponding energy consumption, of Ethernet links \cite{bilal12}. According to ALR, the capacity of each Ethernet link is adjusted, e.g., from 100 Mbps to 1 Gbps or from 1 Gbps to 10 Gbps, to satisfy the incoming traffic while transmitting at the less consuming bit-rate \cite{gunaratne06a,gunaratne06b,gunaratne08}. To date, the widespread of ALR has been slowed due to hardware and implementation issues. The ALR concept can be generalized by the notion of \emph{multi-line rate} (MLR) \cite{idzikowski13a}. %  according to which several discrete capacity values, included the zero-capacity state corresponding to the sleep mode, are available on each link. The analytical study presented in \cite{idzikowski13a} states that MLR approaches are more energy efficient than pure sleep-based ones.

\subsection{Routing Protocol}
Routing protocols have a major impact on traffic engineering techniques for distributing traffic load in the network, and therefore on the energy management polices that can select traffic routes to reduce the consumption or to put in sleep some nodes or links (line cards).

In IP networks, we identify two main classes of routing protocols, i.e., flow based and shortest path based.
When routing is flow based, like in the case of network based on Multi-Protocol Label Switching (MPLS), each traffic demand is routed along one or multiple dedicated paths \cite{wang08}. In this case, traffic engineering is very flexible because routing paths can be selected by the network administrator. However, optimizing the routing paths for each traffic demand can be computationally expensive and require a non negligible  signaling overhead whenever a path update is performed.

When shortest-path routing is adopted \cite{altn09}, like with classical IP destination based forwarding, the shortest paths are determined according to link weights. Therefore, traffic engineering approaches aiming at minimizing energy consumption can be performed by adjusting the link weights. However, since packet forwarding is based on destination address only, not all the combination of paths can be selected and all traffic flows to the same destination must follow the same path(s) regardless of their origins.

From the practical point of view, shortest path routing allows the network administrator to modify only a set of link weights, one for each link, which can be easily updated through state-of-the-art management technologies. However, routing configurations are more constrained and optimization problems become much even more computationally challenging.

Note that both flow based and shortest path routing can be either based on single path or multiple paths (per flow or destination, respectively) .

\subsection{QoS Constraints}
Making IP networks greener should not jeopardize their quality of service (QoS). A first trivial way to avoid congestion and guarantee the QoS is leaving some spare capacity to cope with unpredictable traffic variations and device failures. This is commonly achieved by network operators by imposing a maximum utilization threshold on both network routers and links. 

An alternative strategy (see, e.g., \cite{fortz02}) is to define a convex non-linear function to represent the cost of link/router saturation and to optimize the configuration based on such function.

In case of elastic traffic, where flow rates are dynamically adjusted by congestion control mechanisms, the QoS objectives can be translated into the maximization of utility functions of the rates achievable by each demand \cite{amaldi13c}.

\subsection{Network Survivability and Robustness}
Perfectly tailoring the active capacity to the incoming traffic level is clearly the natural way to maximally reduce the network consumption. However, though costly from the energy perspective, some spare capacity should be always available to cope with unexpected events, including device failures and unexpected traffic variations. There always exists a trade off between energy efficiency and network resilience/robustness.

\subsection{Frequency of network updating}%Optimization Frequency}
Even though  Internet traffic is intrinsically variable, it is typically characterized by a daily/weekly periodic profile \cite{bolla12}, which allows network administrators to estimate the traffic expected in forthcoming days. However, even the most accurate prediction is always affected by some uncertainty. 

Therefore, network engineers must address the following issues: (i) is it better to optimize network configuration in advance according to traffic predictions (planning phase), or in real-time considering direct traffic measurements? (ii) How often must the network be reconfigured to avoid QoS degradation and optimally reduce energy consumption?

The accuracy of off-line approaches is related to the precision of traffic predictions. Assuming that traffic demands are well estimated, off-line methods present
two main advantages. First, the time available to run EANM algorithms is quite large (in the order of hours), allowing the implementation of more complex procedures. Secondly, optimizing in advance allows the careful evaluation of the solutions before their application. However, off-line methods offer scarce flexibility in reacting to unexpected conditions. 

As for on-line approaches which adapt network configuration to real-time measurements, they guarantee prompt reactions to unexpected traffic variations and device failures, but, at the same time drastically reduce the available computational time.

The off-line/on-line nature of the approach also influences the optimization frequency. A common practice for off-line methods is to exploit the slow and periodic dynamics of Internet traffic to identify in advance a few time instances which will trigger pre-determined configuration updates.
Limiting the number of times a network is reconfigured may cause issues in terms of (i) gap between the energy consumed by the currently applied configuration and that requested by the one optimized for the real traffic amount, and (ii) potential inability to meet the requested QoS requirements due to unexpected events. However, recent work have shown that in current IP networks quasi-optimal savings and desired QoS could be achieved through a limited number of network configurations (in terms of routing and device states) to be efficiently applied along an entire day \cite{chiaraviglio11,chiaraviglio13a}.

Differently, with on-line frameworks, the network can be reconfigured every time is needed, but too frequent re-configurations should be avoided to preserve network stability and limit the signalling overhead.

Note that off-line and on-line optimization can be effectively combined to achieve performance optimality as well as network stability and responsiveness to failures.

\subsection{Distributed vs centralized decisions}
EANM can be implemented in a global and coordinated fashion by a centralized entity, e.g., a network management platform, or in a local and distributed manner by means of multiple network agents which are able to locally adjust the configuration and converge towards a stable network configuration.

Centralized approaches manage the entire network as a whole, relying on a set of global measurements which potentially allow to determine quasi-optimal configurations. However, centralized mechanisms are typically computationally complex, require a substantial amount of computing power, and rely on dedicated architectures to collect data and disseminate configuration instructions. 

Conversely, distributed schemes generally exploit limited amounts of local data. They base their strength on speed and simplicity. As expected, these features make very hard to approach solution optimality. It is worth pointing out that centralized EANM is usually implemented off-line, while distributed EANM is generally used for on-line mechanisms.

\section{Energy-Aware Network Optimization Modeling}\label{sec:problems}
The issue of managing the network to reduce its power consumption can be modeled as a mathematical programming problem.
The formulation varies according to the considered network scenario so as to include all the problem features, e.g., routing scheme, energy components, QoS measures.  \emph{Energy Aware Network Management} (EANM) problems are strictly related to another class of optimization problems called \emph{Network Design} (ND). ND shares many modeling features with EANM, and it has been extensively studied in the last fifty years (see, for instance, \cite{MedhiPioro}, \cite{Assad78}, Chapter~"An annotated bibliography in communication network design and routing" of \cite{YuanPhD}).

In this section, we present several mathematical models, starting from the basic EANM problem and then including other network and problem features. We will point out the features common to ND problems, and those which are specific of EANM problems. This tutorial part is meant to assist green networking experts to identify the specific optimization problem which they are interested in or they are working on. We believe that, in both operation research and engineering communities, being always aware of which is the right reference model is crucial to better exploit the existing literature to rigorously compare novel proposals with state of the art.

\subsection{Basic Energy-Aware Network Management Problem}
Let us consider an IP network topology composed by routers and links connecting them. The network can be represented by a directed graph $G=(V,A)$, where the set of nodes $V$ represents the routers and the set of arcs $A$ represents the links. Let $c_{ij} \ge 0$ be the link capacity. 
%Network nodes can be divided in two disjoint sets: the set of core nodes $V^c$, which represents the set of transit routers, and the set of edge nodes $V^e$, which represents the routers that can be both source and destination of traffic flowing through the network.
%\nota{Abbiamo bisogno di distingure nodi di edge e core?}
Traffic is represented by a set of demands $D$. Each traffic demand $d \in D$ is described by a source node $o^d$, a destination node $t^d$ and a non-negative bandwidth request $r^d$.
 
The \emph{IP-Basic Energy-Aware Network Management} (IP-BEANM) problem asks to route the set of demands so as to minimize the network energy consumption.
To represent flow of demand $d$ on each arc $(i,j)$, we introduce non-negative variables $f_{ij}^d$. Additional variables $f_{ij}$ and $f_i$ can be introduced to  represent the total traffic carried by link $(i,j) \in A$ and total traffic entering node $i \in V$, respectively. Note that the flow traversing a node is the sum of the flows entering the node plus the demands originating by such a node. 

Different functions have been used to model network energy consumption. Energy consumption is commonly represented as the sum of the link energy consumption $\Pi_{ij}\left(f_{ij} \right)$ and the node energy consumption $\Pi_i\left(f_i \right)$.  In IP-BEANM such functions are assumed to be continuous and non-decreasing.

The general notation is summarized in Table \ref{tab:elements}.

\begin{table}[!hbt]
\centering
\caption{Elements of the basic problem}\label{tab::parameters}
\vspace{0.2cm}
\begin{tabular}{c|l} \hline \hline
\multicolumn{2}{c}{\bf \textit{Sets}} \\
\hline
%$V^c$   & Core nodes\\
%$V^e$   & Edge nodes\\
$V $     & Network nodes \\
$A \subset V \times V$		& Network links	\\
$D$		& Traffic demands \\
\hline
\hline
\multicolumn{2}{c}{\bf \textit{Parameters}} \\
\hline
$o^d$ & Origin of demand $d$\\
$t^d$ & Destination of demand $d$\\
$r^d$ & Traffic request for demand $d$\\
\hline
%$\pi_i$ & Fixed consumption component for node $i$\\
%$n_{ij}$ & Number of available cards on link $(i,j)$ \\
$c_{ij}$ & Card capacity on link $(i,j)$ \\
%$\mu_{ij}$ & Max link utilization allowed on link $(i,j)$ \\
%$\pi_{ij}$ & Fixed consumption component for link $(i,j)$\\
$\Pi_{ij}\left(f_{ij} \right)$ & Consumption function for link $(i,j)$\\
$\Pi_i\left(f_i \right)$ & Consumption function for node $i$\\
\hline
\hline
\multicolumn{2}{c}{\bf \textit{Variables}} \\
\hline
$f_{ij}^d$ & Amount of flow of demand $d$ and on link $(i,j)$\\
$f_{ij}$& Total amount of flow carried by link $(i,j)$\\
$f_i$	& Total amount of flow carried by node $i$\\
\hline
\hline
\multicolumn{2}{c}{}
\label{tab:elements}
\end{tabular}
\vspace{0.5cm}
\end{table}

The arc based model for IP-BEANM is the following: 
  
\begin{eqnarray}
\label{eq:ob.base}\min \sum_{(i,j)\in A}\Pi_{ij}\left(f_{ij} \right) + \sum_{i\in V}\Pi_{i}\left(f_{i} \right) \\
%\end{eqnarray}
%\vspace{-0.2cm}
%\emph{s.t.}
%\begin{eqnarray}
\label{eq:flow_balance.base}\sum_{\substack{j \in V:\\(i,j)\in A}}f_{ij}^d - \sum_{\substack{j \in V:\\(j,i)\in A}} f_{ji}^d =
\begin{cases}
r^d  & \text{if $i = o^d$} \\
-r^d & \text{if $i = t^d$}\\
0 & \text{otherwise}
\end{cases} & \forall i \in V,\,d \in D\\
\label{eq:total_link_flow.base}\sum_{d\in D} f_{ij}^{d} =  f_{ij} & \forall (i,j) \in A \\
\label{eq:total_node_flow.base}\sum_{(j,i) \in A}f_{ji} + \sum_{\substack{d \in D:\\o^d = i}}r^d =  f_{i} & \forall i \in V \\
%\label{eq:capacity.base}f_{ij} \leq  \mu_{ij}c_{ij} & \forall (i,j)\in A \\
\label{eq:capacity.base} f_{ij} \leq  c_{ij} & \forall (i,j)\in A \\
\label{eq:domain_real.base.arcsdem} f_{ij}^d\geq 0 & \forall (i,j)\in A,\, d \in D\\
\label{eq:domain_real.base.arcs}  f_{ij}\geq 0 & \forall (i,j)\in A\\
\label{eq:domain_real.base.nodes} f_{i} \geq 0 & \forall i \in V.
\end{eqnarray}

Objective function (\ref{eq:ob.base}) minimizes the network power consumption due to flow on arcs and through nodes. Flow conservation constraints \eqref{eq:flow_balance.base} ensure that each traffic demand is routed from its source to its destination, while constraints \eqref{eq:total_link_flow.base} and \eqref{eq:total_node_flow.base} compute the bandwidth requirements for links and nodes, respectively. Link capacity are enforced by constraints (\ref{eq:capacity.base}). Inequalities \eqref{eq:domain_real.base.arcsdem}-\eqref{eq:domain_real.base.nodes} give the variable domain. Different formulations can be taken into account without significantly changing the structure of the problem (for example considering the in-out flow for node capacity and not only in flow).

If only arc energy consumption is considered and functions $\Pi_{ij}(\cdot)$ are linear -- i.e. $\Pi_{ij}=a_{ij}f_{ij}$ --
then the term $\sum_{i\in V}\Pi_{i}\left(f_{i} \right)$ in the objective function disappears, and variables $f_i$ and constraints~\eqref{eq:total_node_flow.base} are no longer necessary and can be discarded. The resulting model is the following:

\begin{eqnarray}
\label{eq:ob.mcmc}\min \sum_{(i,j)\in A}a_{ij}\sum_{d \in D}f_{ij}^d\\
\label{eq:flow_balance.mcmc}\sum_{\substack{j \in V:\\(i,j)\in A}}f_{ij}^d - \sum_{\substack{j \in V:\\(j,i)\in A}} f_{ji}^d =
\begin{cases}
r^d  & \text{if $i = o^d$}, \\
-r^d & \text{if $i = t^d$},\\
0 & \text{otherwise}
\end{cases}, & \forall i \in V\\
\label{eq:total_link_flow.mcmc}\sum_{d\in D} f_{ij}^{d} \le  c_{ij}, & \forall (i,j) \in A \\
\label{eq:domain_real.mcmc}  f_{ij}^d\geq 0, & \forall (i,j)\in A,\, d \in D
\end{eqnarray}

Formulation~\eqref{eq:ob.mcmc}-\eqref{eq:domain_real.mcmc} represents a \emph{multicommodity flow problem} (see Chapter 17 of \cite{Ahuja93}).

The multicommodity flow problem with continuous flow variables can be polynomially solved. The \emph{multicommodity 
integral flow problem}, instead consider integer flow variables and it is NP-complete in its decision form, even if only two commodities and unitary capacities are considered (see Appendix of \cite{Garey79}).

We observe that, in the multicommodity model, node cost or capacity can be accounted for by splitting the nodes. Each node $i$ is replaced by two nodes $i^{'}$, and $i^{''}$, one representing the input side of $i$, the other representing the output side. The arc between $i^{'}$ and $i^{''}$ accounts for node $i$ capacity and cost. The size of the problem will grow to $2N$ nodes and $A+N$ arcs (see Chapter 2 of \cite{Ahuja93}).

Besides linear objective functions, non-linear, both convex and non-convex functions \cite{garroppo13,vasic10,cardonarestrepo09}, are considered in the literature. Piece-wise linear approximations may be applied when convex functions are minimized, (see e.g., \cite{garroppo13,fortz02}), still preserving linearity of the formulation. Convex non-linear functions have also
been used to model other kind of costs/objectives such as delays in the urban networks or congestion in communication networks and such applications are widely addressed in the literature (see Chapter 14 of~\cite{Ahuja93} and~\cite{Florian86}).

\subsubsection{Alternative formulations}\label{sec:perpath}% Alternatives formulations may also be applied to model the problem.

An alternative formulation uses \emph{per-source (or destination) routing scheme} (\cite{pioro04},\cite{amaldi13,chiaraviglio12a}).
Continuous non-negative real variables $f_{ij}^s$ are introduced that represent the amount of traffic carried by link $(i,j)$ and generated by node $s$. Flow conservation constraints \eqref{eq:flow_balance.base} as well as total link flow constraints \eqref{eq:total_link_flow.base} are replaced by:

\begin{eqnarray}
\label{eq:flow_balance.source}\sum_{\substack{j \in V:\\(i,j)\in A}}f_{ij}^s - \sum_{\substack{j \in V:\\(j,i)\in A}} f_{ji}^s =
\begin{cases}
\sum_{d \in D:o^d = s} r^d  & \text{if $i = s$} \\
-r^d & \text{if $\exists d \in D : t^d = i,o^d = s $}\\
0 & \text{otherwise}
\end{cases}, & \quad \forall s \in V,\,i\, \in V\\
\label{eq:total_link_flow.source}\sum_{s\in V} f_{ij}^{s} = f_{ij}, & \forall (i,j) \in A
%\label{eq:domain_real.source}f_{ij}^{s} \geq 0, & \forall s \in V^e,\, (i,j) \in A.
\end{eqnarray}

With \emph{per-source} routing, we pass from $|A||D|$ to $|A||V|$ flow variables and from $|V||D|$ to $|V||V|$  flow conservation constraints, considerably reducing the model size, if the number of demands $|D|$ is larger than the number of nodes $|V|$. It is worth pointing out that the routing of each demand can be easily derived from a feasible solution of the \emph{per-source} model.

Another alternative is to consider \emph{per-path routing}, where the potential routing paths of each demand $d \in D$ are explicitly collected into the set $P^d$.
A fractionary variable $x^{dp} \in \left[0,1 \right]$ is introduced for each demand $d$ that represents the fraction of $r^d$ sent along path $p \in P^d$. Constraints \eqref{eq:flow_balance.base}  and \eqref{eq:total_link_flow.base} are replaced by

\begin{eqnarray}
\label{eq:path_choice.path}\sum_{p \in P^d}x^{dp} = 1, & \forall d \in D, \\
\label{eq:total_link_flow.path}\sum_{\substack{p \in P^e:\\(i,j) \in p}} r^d x^{dp} \leq  f_{ij}, & \forall (i,j) \in A.
\end{eqnarray}

Equations \eqref{eq:path_choice.path} replace flow balancing constraints and ensures that all the incoming traffic $r^d$ is sent along the paths in $P^d$. Constraints \eqref{eq:total_link_flow.path} compute the bandwidth requested on link $(i,j)$.

The \emph{per-path} formulation requires the full description of path set $P^d$ for each demand. However, as the paths are exponentially many, it is often described in an implicit way applying column generation approaches.

\subsection{Unsplittable Routing}\label{sec:sp}
To simplify network management tasks and avoid complex operations such as packet reordering, network administrator prefer to adopt a single path routing scheme. We can observe that simply asking for integer flow variables will not be sufficient to guarantee having a single path for each demand. To obtain single path routing, binary variables $x_{ij}^d \in \{0,1\}$ are introduced, which are equal to one if link $(i,j)$ is used by demand $d$. Furthermore, constraints \eqref{eq:flow_balance.base}-\eqref{eq:total_link_flow.base} must be modified as follows:

\begin{eqnarray}\label{eq:primary_balance.single}
\sum_{\substack{j \in V:\\(i,j)\in A}}x_{ij}^d - \sum_{\substack{j \in V:\\(j,i)\in A}} x_{ji}^d =
\begin{cases}
1  & \text{if $i = o_d$}, \\
-1 & \text{if $i = t_d$},\\
0 & \text{otherwise}
\end{cases} & \forall i \in V,\,d \in D \\
\label{eq:total_link_flow.single}\sum_{d\in D} r^d x_{ij}^{d} =  f_{ij}, & \forall (i,j) \in A
%\label{eq:domain_binary.single}x_{ij}^d \in \{0,1\}, & \forall (i,j) \in A,\, d \in D. 
\end{eqnarray}

Constraints (\ref{eq:primary_balance.single}) replace flow balancing constraints, while equations \eqref{eq:total_link_flow.single} determine the total flow carried by each arc.
We can observe that for the per-path routing, asking variables $x^{dp}$ to be binary is sufficient to obtain a single path routing. On the contrary, per-source (per-destination) routing are not suitable to model unsplittable demands.

\subsection{Congestion reduction constraint}
Network congestion is usually prevented by the network operator by means of a maximum link utilization threshold $\mu_{ij}$. In this case capacity constraints~\eqref{eq:capacity.base} will become:
\begin{eqnarray}
\label{eq:capacity.cong}f_{ij} \leq  \mu_{ij}c_{ij} & \forall (i,j)\in A
\end{eqnarray}
We can observe that this modification does not change the structure of the problem, in fact simply scaling by the coefficient $\mu_{ij}$ the capacity of arc $(i,j)$ will result in the same constraint of IP-BEANM.  From now on we will consider the constraint with congestion reduction mechanism, in fact, when no congestion reduction is considered it can be obtained setting $\mu_{ij}=1$.

\subsection{Sleeping capabilities}\label{sec:sleeping}
The consumption profile of current network devices is quite well approximated by an ON-OFF (step) curve (\cite{chabarek08,mahadevan09,mellah09,vanheddeghem12b}): that means that the static consumption component, i.e., the energy required to keep a device on, represents more than 90\% of the total power required by a fully used device.  Indeed, most papers on energy-aware network management ignore utilization-based energy costs and considers only a fixed consumption $\pi_i$ for each active node $i \in V$ and $\pi_{ij}$ for each active link $(i,j) \in A$. The energy footprint of the network is thus reduced by putting to sleep unused network devices.

To model the power state of both routers and links, binary variables $y_i$ and $w_{ij}$ are introduced which are equal to 1 if node $i \in V$ and link $(i,j) \in A$, respectively, are powered on. The energy consumption to be minimized is thus:

\begin{eqnarray}
\label{eq:ob.sleep} \sum_{(i,j)\in A}\pi_{ij}w_{ij} + \sum_{i\in V}\pi_{i}y_i
\end{eqnarray}

Capacity constraints (\ref{eq:capacity.base}) must be modified and additional constraints must be included:
\begin{eqnarray}
\label{eq:capacity.sleep}f_{ij} \leq  \mu_{ij}c_{ij}w_{ij} & \forall (i,j)\in A,\\
\label{eq:coherence1.sleep} w_{ij} \leq y_{i}, & \forall (i,j) \in A \\
\label{eq:coherence2.sleep} w_{ij} \leq y_{j}, & \forall (i,j) \in A
%\label{eq:domain_real.sleep1}y_i \in \{0,1\}, & \forall i \in V\\
%\label{eq:domain_real.sleep2}w_{ij} \in \{0,1\}, & \forall (i,j) \in A.
\end{eqnarray}

Constraints~\eqref{eq:capacity.sleep} prevent routing a demand through a sleeping link, and replace constraints~\eqref{eq:capacity.base}. Constraints~\eqref{eq:coherence1.sleep}-\eqref{eq:coherence2.sleep} forbid the activation of a link connected to a sleeping router.
If $\pi_i =0$, the second term of the objective function can be neglected and all  routers are assumed to be always active: constraints \eqref{eq:coherence1.sleep} and \eqref{eq:coherence2.sleep} can be neglected , as well. The resulting model shares the structure with the Network Design problem, where a minimum cost subset of arcs must be selected (see~\cite{MedhiPioro}).

\subsection{Bundled Links, Adaptive Link Rate and Multi-Line rate}

\subsubsection{Bundled links}\label{sec:bundled}

A single network link $(i,j) \in A$ can be seen as a bundle of $n_{ij}$ line cards (or cables) of the same type, namely having the same capacity and same power consumption. Each line card can be switched on or off independently from the others, so as to reduce the energy consumption if less than $n_{ij}$ cards are needed. Binary variables $w_{ij}$ are not suitable for describing the possible statuses of a link, thus, when bundled links are considered, $w_{ij}$ are integer non-negative variables: $w_{ij} \in \left[  0,\ldots,n_{ij}\right]$, representing the number of active line cards on each link. Furthermore, constraints~\eqref{eq:coherence1.sleep} and \eqref{eq:coherence2.sleep} should be slightly modified:

\begin{eqnarray}
\label{eq:coherence1.sleepbundled} w_{ij} \leq n_{ij}y_{i}, & \forall (i,j) \in A \\
\label{eq:coherence2.sleepbundled} w_{ij} \leq n_{ij}y_{j}, & \forall (i,j) \in A
%\label{eq:domain_integer.bundled}w_{ij} \in \left[0,\ldots,n_{ij}\right], & \forall (i,j) \in A.
\end{eqnarray}
 
The Network Loading problem with modular capacity, namely the problem of allocating the minimum cost number of modular capacity devices on network arcs, has  a similar model structure (see~\cite{MedhiPioro}).

\subsubsection{Adaptive Link Rate (ALR) and Multi-Line rate (MLR)}\label{sec:alr}
According to ALR or MLR, each link can work at different rates with specific speed and energy consumption. For instance, Ethernet links can be operated at four different rates, i.e., 10 Mbps, 100 Mbps, 1 Gbps and 10 Gbps, each one with a particular consumption requirement: switching from 100 Mbps to 1 Gbps requires 4 additional Watts, while for 10 Gbps links the increment is around 15 W \cite{gunaratne08}. MLR can be similarly implemented by means of the so-called channel bonding \cite{hahnel13}, according to which a virtual link is mapped over multiple physical links which, in this energy-aware case, can be activated only one at a time.

The set $E_{ij}$ containing the different working configurations $e$ is introduced for each link $(i,j) \in A$. Each configuration $e \in E_{ij}$ is described by a capacity $c_{ij}^e$ and a fixed consumption, $\pi_{ij}^e$. Note that link sleeping status is represented by a configuration with zero capacity and zero consumption. Such configuration does not request the activation of the end points.
Furthermore, link state variables $w_{ij}$ must be replaced by variables $w_{ij}^e$ which are equal to 1 if link $(i,j)$ is operated with configuration $e$.

The objective accounts for the power of the configuration used by each link:
\begin{eqnarray}
\label{eq:ob.alr}\min \sum_{(i,j)\in A}\sum_{e \in E_{ij}}\pi_{ij}^e w_{ij}^e + \sum_{i\in V}\pi_{i}y_i,
\end{eqnarray}

Constraints~\eqref{eq:capacity.sleep}-\eqref{eq:coherence2.sleep} must be restated and configuration selection constraints must be added:
\begin{eqnarray}
\label{eq:bonding.alr}\sum_{e \in E_{ij}}w_{ij}^e = 1, & \forall (i,j)\in A \\
\label{eq:capacity.alr}f_{ij} \leq  \mu_{ij}\sum_{e \in E_{ij}}c_{ij}^e w_{ij}^e, & \forall (i,j)\in A \\
\label{eq:coherence1.alr}\sum_{\substack{e \in E_{ij}:\\c_{ij}^e \neq 0}} w_{ij}^e \leq y_{i}, & \forall (i,j) \in A \\
\label{eq:coherence2.alr}\sum_{\substack{e \in E_{ij}:\\c_{ij}^e \neq 0}} w_{ij}^e \leq y_{j}, & \forall (i,j) \in A
%\label{eq:domain_binary.alr}w^e_{ij} \in \{0,1\}, & (i,j) \in A,\, e \in E_{ij}. 
\end{eqnarray}

Constraints~\eqref{eq:bonding.alr} ensure that only one configuration is selected for each link, while constraints~\eqref{eq:capacity.alr} guarantee that link maximum utilization threshold of the chosen link configuration is not exceed. Finally, constraints~\eqref{eq:coherence1.alr}-\eqref{eq:coherence2.alr}, which are a restated version of the original \eqref{eq:coherence1.sleep}-\eqref{eq:coherence2.sleep}, allow to select a non-zero capacity configuration for a link $(i,j) \in A$ only if the chassis are active on both sides of link $(i,j)$ itself.

\subsection{Protection from single link failures}\label{sec:pr}
The energy consumption of communications networks is typically optimized by putting to sleep the redundant resources at both link and node levels, as mentioned before. However, the need for greener networks should not compromise the capability of the networks to react to unexpected events such as device failures. Failure events are usually handled by automatic protection mechanisms that redirect the traffic over still working routes. The most popular strategy is to reserve backup bandwidth.  Being link failures, and even more node failures, a very unlikely event, the majority of protection mechanisms is focused on managing single link failures.

The most straightforward form of protection is known as point-to-point protection, where a backup path is computed for each demand, independent from the fault. Primary and backup paths must be link disjoint. This protection strategy can be implemented in networks operated with flow-based routing protocol, e.g., MPLS, and configured to allow single path routing only.

To model point-to-point protection,  backup paths variable and flow conservation constraints are needed. 
Let $\xi_{ij}^d$ be the binary variables which are equal to 1 if link $(i,j) \in A$ is used by the backup path of demand $d \in D$. The additional constraints to represent the protection scheme are:
 
\begin{eqnarray}\label{eq:backup_balance.pr}
\sum_{\substack{j \in V:\\(i,j)\in A}}\xi_{ij}^d - \sum_{\substack{j \in V:\\(j,i)\in A}} \xi_{ji}^d =
\begin{cases}
1  & \text{if $i = o_d$}, \\
-1 & \text{if $i = t_d$},\\
0 & \text{otherwise},
\end{cases}  & \forall i \in V,\,d \in D \\
\label{eq:disj1.pr}x_{ij}^{d} + \xi_{ij}^{d} \leq 1, & \forall (i,j) \in A,\, d \in D \\
\label{eq:disj2.pr}x_{ij}^{d} + \xi_{ji}^{d} \leq 1, & \forall (i,j) \in A,\, d \in D. 
\end{eqnarray}

 Constraint~\eqref{eq:backup_balance.pr} are the flow conservation for the backup paths, while \eqref{eq:disj1.pr}-\eqref{eq:disj2.pr} guarantee that primary and backup paths are link disjoint.

Furthermore, it is necessary to modify capacity constraints to account also for backup flows. The way backup flow is computed differentiates different kinds of end-to-end protection schemes. The most used are the dedicated and the shared one. 
The \emph{dedicated} protection scheme reserves the same amount of bandwidth on each link of the primary and the backup paths. The dimensioning constraint~\eqref{eq:total_link_flow.single} is replace by:
\begin{eqnarray}
\label{eq:total_link_flow.pr}
\sum_{d \in D} r^d \left(x_{ij}^d +\xi_{ij}^d\right)=f_{ij} & \forall (i,j) \in A
\end{eqnarray}

which accounts both for the primary and the backup flows.

As simultaneous link failures are assumed to never occur, \emph{shared} protection aims at reducing the reserved backup capacity by allowing backup paths to share  resources, if their primary paths are failure disjoint. In fact, under the single link failure assumption these backup paths will never be activated simultaneously. Therefore, different link failures require different amounts of capacity for each link. The backup capacity to be reserved for each link is forced by the failure scenario which requires the highest possible backup capacity, i.e. the failure that affects the highest amount of demands.

To correctly compute the backup bandwidth reserved on each link, binary variables $g_{ijkl}^{d}$ are introduced, which are equal to 1 if traffic demand $d \in D$ is rerouted on link $(i,j) \in A$ when a failure occurs on link $(k,l) \in A$ (i.e. if traffic demand $d \in D$ uses a primary and a backup paths routed, respectively, on link $(i,j) \in A$ and link $(k,l) \in A$).
The following constraints replace (\ref{eq:total_link_flow.pr}) to respect the link maximum utilization:

\begin{eqnarray}
\label{eq:bottleneck.pr} g_{ijkl}^d \geq x_{ij}^d+ \xi_{kl}^d - 1, & \forall (i,j),(k,l) \in A,\, d \in D,\\
\label{eq:shared_link_cap.pr}\sum_{d \in D}r^d\left(x_{ij}^d+g_{klij}^d\right)=f_{ij} & \forall (i,j),(k,l) \in A. 
\end{eqnarray}

\subsection{Multi-Periodicity}\label{sec:mp}
%Qui direi che dobbiamo spiegare il fatto che il problema si moltiplica sul set di periodi S. Possiamo anche presentare i vincoli inter-periodo che noi usiamo (sia i reliability constraint che quelli per i costi delle accensioni).

In IP networks, traffic conditions are not constant but vary throughout the day. Accordingly, the network configuration can be adjusted to accommodate the incoming level of traffic. Each change of configuration may result in an additional energy consumption due to reactivation of sleeping elements. Furthermore, the number of changes can be limited for instance by constraining the changes of routing paths or the number of reactivation for a specific device, in fact too frequent state switching could negatively affect the device lifetime. The resulting multiperiod problem consists in planning network configuration along a given time horizon, so as to minimize the overall network energy consumption. As shown in some recent studies \cite{chiaraviglio11,chiaraviglio13a}, due to the slow dynamic of Internet traffic, a few time periods with a duration in the order of hours are enough to provide a good representation of traffic.

The former models can be extended to take into account multiperiodicity. Let $S$ denotes the set of time periods in the considered time horizon. Demand varies in different time periods: let $r^{d\sigma}$ denote the amount of traffic of demand $d$ in time period $\sigma$. As the traffic variations are usually cyclic,  the multiperiod problem must account also for the transition between the last and the first period. Single period path variables $x_{ij}^{d}$ are replaced by the variables $x_{ij}^{d\sigma}$, which are equal to 1 if link $(i,j) \in A$ is used to route demand $d \in D$ during period $\sigma \in S$. All the other variables are changed accordingly. Routing constraints should be defined for each single time period:

\begin{eqnarray}
\label{eq:flow_balance.multi}\sum_{\substack{j \in V:\\(i,j)\in A}}x_{ij}^{d\sigma} - \sum_{\substack{j \in V:\\(j,i)\in A}} x_{ji}^{d\sigma} =
\begin{cases}
1  & \text{if $i = o^d$} \\
-1 & \text{if $i = t^d$}\\
0 & \text{otherwise}
\end{cases} & \forall i \in V,\,d \in D,\, s \in S\\
\label{eq:total_link_flow.base_multi}\sum_{d\in D} r^{d\sigma}x_{ij}^{d\sigma} =  f_{ij}^{\sigma} & \forall (i,j) \in A,\, \forall s \in S\\
\end{eqnarray}

The optimal solution of the multi-period problem cannot be reached by treating independently each time period, as inter-period constraints are considered, which account for energy consumption due device reactivation, limit in the number of reactivations, or in the number of changes in routing paths.
 
Let non-negative real variables $z_j^{\sigma}$ denote the energy consumed by chassis $j \in V$ if it is  activated at the beginning of period $\sigma \in S$ and binary variables $u_{ijk}^{\sigma}$ denote the activation state of line card $k$: they are equal to 1 if the $k$-th line card of link $(i,j) \in A$ is activated from the sleeping state at the beginning of period $\sigma \in S$. 

The value of variables $z$ is computed accordingly to the following constraints:

\begin{eqnarray}
\label{eq:reactivation_power.multiper}z_j^{\sigma} \geq \delta \pi_j \left(y_{j}^{\sigma} - y_{j}^{\sigma-1} \right), & \forall j \in V,\, \sigma \in S,
\end{eqnarray}

\noindent where $\delta$ is a parameter in $\left[0,1\right]$ representing the additional fraction of the nominal chassis consumption which is consumed to reactivate chassis $j \in V$.

Being $\eta_{on}$ the maximum number of reactivations allowed to each line card along the whole time-horizon, the switching limit is imposed by means of:

\begin{eqnarray}
\label{eq:card_state1.multiper}\sum_{k=1}^{n_{ij}} u_{ijk}^\sigma \geq w_{ij}^{\sigma} - w_{ij}^{\sigma-1},& \forall (i,j) \in A,\, \sigma \in S \\
\label{eq:card_state2.multiper}\sum_{\sigma \in S} u_{ijk}^\sigma \leq \eta_{on},& \forall (i,j) \in A,\, k \in \left\lbrace 1,\ldots, n_{ij}\right\rbrace .
\end{eqnarray}

Eq. \eqref{eq:card_state1.multiper} forces the model to reactivate the required number of line cards at the beginning of each time period, while Eq. \eqref{eq:card_state2.multiper} impose the reactivation limit for each single line card.

Finally, let us consider the path adjustment limitation. To use the same paths along the whole time horizon we can consider the same flow or path variables (according to which variables are required by the single period problem) for all the time periods keep the original flow conservation constraints (they are not defined for all $\sigma \in S$).

\subsection{Shortest path routing models}
When network routing follows a shortest path scheme such as, for instance, in IP networks operated with the very popular OSPF routing protocol, the path used by each demand cannot be explicitly selected, but it is indirectly determined by the set of administrative weights manually assigned by the network administrator. This scheme, which, from a practical perspective, scales well with respect to the number of traffic demands (the number of link weights is always the same), is typically exploited to route best effort traffic. 

To favor load balancing and prevent network congestion, shortest path protocols are usually instructed to respect the so-called \emph{Equal Cost Multi-Path} (ECMP) rule, according to which, at each node, a certain traffic demand is equally split among all the outgoing line cards belonging to at least one shortest path toward the corresponding destination.

When routing is shortest-path, performing traffic engineering, i.e., optimizing the routing paths, is equivalent to adjusting the link weights. To model link weight optimization new variables and constraints must be introduced.  Let $u_{ij}^{t}$ be the binary variables which are equal to 1 if link $(i,j) \in A$ lies on a shortest path between $i$ and node $t$, let $z_{i}^{t}$ bet the non-negative real variables which represent the amount of bandwidth reserved on all the outgoing links of $i$ which belong to at least a shortest path from $i$ to $t$. Finally, let $l_{i}^{t}$ be the non-negative real variables accounting for the minimum distance between $i$ and $t$, and let $\omega_{ij}$ be the non-negative real variables which report the link weight of link $(i,j) \in A$. Being $M$ a large enough constant, shortest path routing is modeled by adding the following constraints \cite{pioro04,amaldi13}

\begin{eqnarray}
\label{eq:commonflow.sp} 0 \leq z_{i}^{t}-\sum_{\substack{d \in D:\\t_d = t}}f_{ij}^{d} \leq (1-u_{ij}^{t})\sum_{\substack{d \in D:\\t_d = t}}r^d, & \forall t\in V,\, (i,j) \in A \\
\label{eq:noflow_nosp.sp} \sum_{\substack{d \in D:\\t_d = t}}f_{ij}^{d} \leq u_{ij}^{t}\sum_{\substack{d \in D:\\t_d = t}}r^d, & \forall t\in V,\,(i,j)\in A
\end{eqnarray}
\begin{eqnarray}
\label{eq:rightweight1.sp}0\leq l_{j}^{t}+\omega_{ij}-l_{i}^t\leq(1-u_{ij}^{t})M,&\forall t\in V,\,(i,j)\in A\\
\label{eq:rightweight2.sp}1-u_{ij}^{t} \leq l_{j}^{t}+\omega_{ij} - l_{i}^{t}, \forall t\in V,\,(i,j)\in A\\
\label{eq:rightweight3.sp}u_{ij}^t \leq w_{ij}, & \forall t \in V,\,(i,j)\in A \\
\label{eq:maxweight.sp}\omega_{ij} \geq (1-w_{ij})\omega_{max}, & \forall(i,j)\in A \\
\label{eq:weightvalues.sp}1 \leq \omega_{ij} \leq \omega_{max}, & \forall(i,j)\in A \\
\label{eq:domain_binary.sp}u_{ij}^{t} \in \{0,1\}, & \forall t\in V,\, (i,j)\in A \\
\label{eq:domain_real.sp} l_{h}^t,\,z_h^t,\,\omega_{ij} \geq 0, & \forall h,t \in V,\, (i,j) \in A.
\end{eqnarray}

Constraints~\eqref{eq:commonflow.sp}-\eqref{eq:noflow_nosp.sp} ensure that the ECMP rule is respected by correctly splitting among the outgoing interfaces the traffic which crosses each node: that means, (i) the whole traffic destined to node $t \in V$ is equally split among all the outgoing interfaces belonging to at least one shortest path toward $t$, and (ii) no traffic is routed on the remaining line cards. Shortest path constraints \eqref{eq:rightweight1.sp}-\eqref{eq:rightweight2.sp} define the relationship between shortest paths and link weights, while constraints~\eqref{eq:rightweight3.sp}-\ref{eq:weightvalues.sp} are used to correctly configure the weight of the sleeping links. Constraints~\eqref{eq:rightweight3.sp} state that a sleeping link cannot belong to any shortest path, constraints~\eqref{eq:maxweight.sp} force the weight of a sleeping link to assume the maximum vale $\omega_{max}$, and constraints~\eqref{eq:weightvalues.sp} define the existence domain of link weight variables. Finally the domain of the remaining variables is expressed by constrains~\eqref{eq:domain_binary.sp}-\eqref{eq:domain_real.sp}.

Shortest path constraints introduce significant complexity which makes the resulting traffic engineering problems hardly tractable by classic exact approaches (both energy-unaware \cite{altn09} and energy-aware \cite{amaldi13} problems). Therefore, heuristic methods are naturally preferred in literature to handle this class of problems. For the sake of completeness, we refer the reader to \cite{lee12b} for a non-linear mixed-integer formulation which addresses the energy-aware traffic engineering problem with shortest path routing by considering a \emph{per-path} approach.

%%\nota{Qui bisogna riportare tutto quello che si puo' e si e' applicato nel caso shortest path routing}

\subsection{Other peculiar features}%Questa parte riguarda la parte per la gestione delle domande elastiche. Dobbiamo decidere quanto a lungo e approfonditamente parlarne.
\subsubsection{Other forms of protection}
If needed, other survivability features can be modelled, such as multiple failures, Shared risk groups, or restoration mechanisms. All such features have been addressed in the Network Design literature. In particular, the smart protection paradigm has been proposed for accounting for energy consumption in case of failure. The failures are a very unlikely event and the failed links are reasonably assumed to be promptly restored. Therefore a link hosting only backup paths is needed only for short time and its consumption is negligible compared with the one of links hosting primary paths. Thus, the objective energy function depends only on primary paths flows.

\subsubsection{Elastic traffic}\label{sec:elastic}
Although traffic demands are typically considered as inelastic, i.e., with a fixed transmission rate, it is worth pointing out that all the traffic flow carried by the very popular TCP protocol are by definition elastic. That means that the rate of a given flow is dynamically determined by the distributed congestion control mechanisms of TCP (or other technologies) to maximally exploit the capacity available on the routing path. When multiple flows share the same network resources, the role of the congestion control mechanisms is to distribute the contended resources in a fair manner among the concurrent demands. In case of elastic traffic, the network administrator has no control on the flow rates, which are uniquely determined by the congestion control mechanism according to the routing paths chosen by the administrator itself.

\subsubsection{Robustness}
Another important aspect concerns the intrinsic uncertainty which affects traffic demands \cite{addis13,coudert13}. Classic techniques for Robust Optimization in ILP flow problems can be naturally adapted to account for traffic variation in the energy-aware network management scenario. In particular, we refer the reader to the cardinality constrained method proposed by Bertsimas et al. \cite{bertsimas11}, which considers uncertain traffic demands varying within a close symmetric interval centered around their nominal expected values. %This method has been also applied in an energy-aware context \cite{addis13,coudert13}. 

\section{Literature review}
\label{sec:review}
The classifications presented in Sections \ref{sec:approaches} and \ref{sec:problems} is here exploited to present a detailed literature review on EANM approaches. When not explicitly declared in the reference paper, we personally derive the optimization problem solved by each discussed procedure.  Proposals are split on two major families, i.e., flow-based routing and shortest part routing strategies. 
In what follows, we first describe how the different articles were classified. Next, we enter in-depth into the main features of
each set of articles, depending on whether they belong to the flow-based, shortest path or ``other" routing strategies. 

\subsection{Literature classification}

The classification of the vast literature review, is done on two axes. First, for each of the two major routing strategies, the articles are classified according to the problem features,
as presented in Section  \ref{sec:problems}.
Tables \ref {tab:problem_features_fb} and \ref {tab:problem_features_sp} represent that classification.
The tables contain the referenced papers, and entries stating whether the problem treated considered on-off nodes, bundle links, elastic consumption, special QoS constraints, protection or survivability or (un)splittable routing. The tables are concluded with special notes concerning the proposals presented. 

Observing the two tables, one can see, above all, that there are much more articles that deal with flow-based routing and that among them, the large majority deals with
some form of ALR and with unsplittable routing. Some work deal with elastic consumption  and surprisingly few consider multi period, survivability or special QoS constraints.

\begin{landscape}
\begin{table}[tbp]
\centering

\begin{scriptsize}

\tabcolsep 3pt
\renewcommand\arraystretch{1.1}
\caption{Literature classification according to problem features: flow-based routing}
\begin{tabular}{cccccccccc}
\multirow{2}{*}{{\bf Problem}} & \multicolumn{1}{c}{\multirow{2}{1cm}{\centering Nodes off}} & \multicolumn{1}{c}{\multirow{2}{2cm}{\centering MLR, ALR, Bundled Links}} & \multicolumn{1}{c}{\multirow{2}{1.5cm}{\centering Elastic consumpt.}} & \multicolumn{1}{c}{\multirow{2}{2cm}{\centering Multiperiod}} & \multicolumn{1}{c}{\multirow{2}{2cm}{\centering Special QoS constraints}} & \multicolumn{1}{c}{\multirow{2}{1.3cm}{\centering Traffic Unaware}} & \multicolumn{1}{c}{\multirow{2}{2cm}{\centering Protection, Survivability}} & \multicolumn{1}{c}{\multirow{2}{1.5cm}{\centering Unsplittable routing}} & \multicolumn{1}{c}{\multirow{2}{1.5cm}{\centering Special features}} \\
\\ 
\hline 
{\bf Work} & & & & & & & & & \\
%\multicolumn{10}{c}{\multirow{2}{*}{\centering \large \bf \emph{Flow based routing}}} \\ 
%\\
\cite{addis14} & x & x &  & x &  &  &  &  x &  \\ 
\hline
\cite{addis12b} & x & x &  & x &  &  &  & x &  \\ 
\hline
\cite{addis12,addis12a} & x & x &  & x &  &  & x & x &  Smart protection\\ 
\hline
\cite{addis13} & x & x &  & x &  &  &  & x & Robustness  \\ 
\hline
\cite{addis14a} & x & x &  & x &  &  & x & x & Robustness, Smart protection \\ 
\hline
\cite{amaldi13c} & x &  &  &  &  &  &  & x & Elastic traffic \\ 
\hline
\cite{sanso09} & x & & x & & & & & & Bi-objective function\\
\hline
\cite{chabarek08} & x &  &  &  &  &  &  &  & Network design decisions\\ 
\hline
\cite{zhang10} &  & x &  &  &  &  &  &  & \\ 
\hline
\cite{chiaraviglio12a,chiaraviglio08,chiaraviglio09,chiaraviglio09a,lee12a,cheung14} & x &  &  &  &  &  &  &  &  \\ 
\hline
\multicolumn{1}{c}{\multirow{2}{*}{\centering \cite{garroppo13a,garroppo13}}} & \multicolumn{1}{c}{\multirow{2}{*}{\centering x}}  & \multicolumn{1}{c}{\multirow{2}{*}{\centering x}} & \multicolumn{1}{c}{\multirow{2}{*}{\centering x}}  & \multicolumn{1}{c}{\multirow{2}{*}{}}  & \multicolumn{1}{c}{\multirow{2}{*}{}}  & \multicolumn{1}{c}{\multirow{2}{*}{}}  & \multicolumn{1}{c}{\multirow{2}{*}{}}  & \multicolumn{1}{c}{\multirow{2}{*}{\centering }}  & \multicolumn{1}{c}{\multirow{2}{2.5cm}{\centering Piece-wise concave energy costs}} \\
\\
\hline 
\cite{garroppo11} & x &  & x &  &  &  &  &  & Bidirectional link capacity \\ 
\hline
\cite{garroppo12} &  & x & x &  &  &  &  &  &  \\ 
\hline
\cite{giroire10} &  &  &  &  &  &  &  &  & \\ 
\hline
\cite{fisher10} &  & x &  &  &  &  &  &  &  \\ 
\hline
\cite{lin13} &  & x &  &  &  &  &  & x &  \\ 
\hline
\cite{lin12,lin14} & x & x &  &  &  &  &  &  &  \\ 
\hline
\cite{mumey12,wu13} & x & x &  &  &  &  &  &  &  \\ 
\hline
\cite{charalambides13} &  & x &  &  &  &  &  &  & Multi-topology routing\\ 
\hline
\cite{galan-jimenez13a,galan-jimenez13b} &  & x &  &  &  &  &  &  & Energy-Levels \\ 
\hline
\multicolumn{1}{c}{\multirow{2}{*}{\centering \cite{kist11,cianfrani12a,coiro13a,cianfrani13}}} & \multicolumn{1}{c}{\multirow{2}{*}{\centering x}} & \multicolumn{1}{c}{\multirow{2}{*}{\centering }} & \multicolumn{1}{c}{\multirow{2}{*}{\centering }} & \multicolumn{1}{c}{\multirow{2}{*}{\centering }} & \multicolumn{1}{c}{\multirow{2}{*}{\centering }} & \multicolumn{1}{c}{\multirow{2}{*}{\centering }} & \multicolumn{1}{c}{\multirow{2}{*}{\centering }} & \multicolumn{1}{c}{\multirow{2}{*}{\centering }} & \multicolumn{1}{c}{\multirow{2}{4cm}{\centering Node expansion, Multiple sleeping states}} \\
\\ 
\hline
\cite{vasic10,vasic11} &  & x & x &  &  &  &  &  &  \\ 
\hline
%\multicolumn{1}{c}{\multirow{2}{*}{\centering \cite{avallone12}}} & \multicolumn{1}{c}{\multirow{2}{*}{\centering x}} & \multicolumn{1}{c}{\multirow{2}{*}{\centering }} & \multicolumn{1}{c}{\multirow{2}{*}{\centering }} & \multicolumn{1}{c}{\multirow{2}{*}{\centering }} & \multicolumn{1}{c}{\multirow{2}{*}{\centering x}} & \multicolumn{1}{c}{\multirow{2}{*}{\centering }} & \multicolumn{1}{c}{\multirow{2}{*}{\centering }} & \multicolumn{1}{c}{\multirow{2}{*}{\centering }} & \multicolumn{1}{c}{\multirow{2}{4cm}{\centering Admission control and traffic engineering}} \\ 
\cite{avallone12} & x &  & &  & x &  & & x &  \\ 
\hline
\cite{seoane11} &  &  & x &  &  &  &  & x & \\ 
\hline
\cite{kim11,kim12} & x &  & x &  & x & x &  &  x& \\ 
\hline
\cite{coiro13} &  &  &  &  &  &  &  & x &  \\ 
\hline
%\multicolumn{1}{c}{\multirow{2}{*}{\centering \cite{takeshita12}}} & \multicolumn{1}{c}{\multirow{2}{*}{\centering x}} & \multicolumn{1}{c}{\multirow{2}{*}{\centering x}} & \multicolumn{1}{c}{\multirow{2}{*}{\centering x (for nodes)}} & \multicolumn{1}{c}{\multirow{2}{*}{\centering }} & \multicolumn{1}{c}{\multirow{2}{*}{\centering x}} & \multicolumn{1}{c}{\multirow{2}{*}{\centering }}& \multicolumn{1}{c}{\multirow{2}{*}{\centering }} & \multicolumn{1}{c}{\multirow{2}{*}{\centering }} & \multicolumn{1}{c}{\multirow{2}{4cm}{\centering Brute force generation of all the possible power states}}  \\ 
\cite{takeshita12} & x &  & x &  & x & & & x & \\
\hline
\multicolumn{1}{c}{\multirow{2}{*}{\centering \cite{hou14}}} & \multicolumn{1}{c}{\multirow{2}{*}{\centering }} & \multicolumn{1}{c}{\multirow{2}{*}{\centering }} & \multicolumn{1}{c}{\multirow{2}{*}{\centering x}} & \multicolumn{1}{c}{\multirow{2}{*}{\centering }} & \multicolumn{1}{c}{\multirow{2}{*}{\centering x}} & \multicolumn{1}{c}{\multirow{2}{*}{\centering }} & \multicolumn{1}{c}{\multirow{2}{*}{\centering }}  & \multicolumn{1}{c}{\multirow{2}{*}{\centering }} & \multicolumn{1}{c}{\multirow{2}{4cm}{\centering Bi-objective function for energy and delay}} \\
\\ 
\hline
\cite{yang15} & & x &  & x  &  &  & x & x & hop-by-hop routing \\ 
\hline
\cite{francois12,francois13} & x &  &  & x  &  &  & x & x & Two macro-periods \\ 
\hline
\cite{francois14} &  & x &  &  &  &  &  & x &  \\ 
\hline
\cite{bianzino12a} &  &  &  &  &  &  &  &  & \\ 
\hline
\cite{zhao13} &  &  & x &  &  &  &  &  & \\ 
\hline
\cite{giroire12,koster13,coudert13} & x &  &  &  &  &  &  &  & Redundancy elimination \\ 
\hline
\cite{niewiadomska-szynkiewicz13,niewiadomska14} & x & x & x &  &  &  &  &  x&  \\ 
\hline
\multicolumn{1}{c}{\multirow{2}{*}{\centering \cite{luo13}}} & \multicolumn{1}{c}{\multirow{2}{*}{\centering x}} & \multicolumn{1}{c}{\multirow{2}{*}{\centering x}} & \multicolumn{1}{c}{\multirow{2}{*}{\centering }}  & \multicolumn{1}{c}{\multirow{2}{*}{\centering }}  & \multicolumn{1}{c}{\multirow{2}{*}{\centering }}  & \multicolumn{1}{c}{\multirow{2}{*}{\centering }}  & \multicolumn{1}{c}{\multirow{2}{*}{\centering x}} & \multicolumn{1}{c}{\multirow{2}{*}{\centering }} & \multicolumn{1}{c}{\multirow{2}{4cm}{\centering Approximated shared and dedicated protection}} \\
\\ 
\hline
\cite{aldraho12} & x &  & x &  &  &  & x & x &  \\ 
\hline
\cite{lin15} & & x & & & &  &x & & Failure probabilities\\
\hline
\end{tabular}
\label{tab:problem_features_fb}

\end{scriptsize}
\end{table}
\end{landscape}

\begin{landscape}
\begin{table}[tbp]
\centering

\begin{scriptsize}

\tabcolsep 3 pt
\renewcommand\arraystretch{1.1}
\caption{Literature classification according to problem features: shortest path routing}
\begin{tabular}{cccccccccc}
\multirow{2}{*}{{\bf Problem}} & \multicolumn{1}{c}{\multirow{2}{1cm}{\centering Nodes off}} & \multicolumn{1}{c}{\multirow{2}{2cm}{\centering MLR, ALR, Bundled Links}} & \multicolumn{1}{c}{\multirow{2}{1.5cm}{\centering Elastic consumpt.}} & \multicolumn{1}{c}{\multirow{2}{2cm}{\centering Multiperiod}} & \multicolumn{1}{c}{\multirow{2}{2cm}{\centering Special QoS constraints}} & \multicolumn{1}{c}{\multirow{2}{1.3cm}{\centering Traffic Unaware}} & \multicolumn{1}{c}{\multirow{2}{2cm}{\centering Protection, Survivability}} & \multicolumn{1}{c}{\multirow{2}{1.7cm}{\centering Unsplittable routing}} & \multicolumn{1}{c}{\multirow{2}{1.7cm}{\centering Special notes}} \\
\\ 
\hline
{\bf Work} & & & & & & & & & \\ 
%\multicolumn{10}{c}{\multirow{2}{*}{\centering \large \bf \emph{Shortest path routing}}} \\ 
%\\
\multicolumn{1}{c}{\multirow{2}{*}{\centering \cite{amaldi13,amaldi11,amaldi11a,capone13}}} & \multicolumn{1}{c}{\multirow{2}{*}{\centering x}} & \multicolumn{1}{c}{\multirow{2}{*}{\centering }} & \multicolumn{1}{c}{\multirow{2}{*}{\centering }} & \multicolumn{1}{c}{\multirow{2}{*}{\centering }} & \multicolumn{1}{c}{\multirow{2}{*}{\centering x}} & \multicolumn{1}{c}{\multirow{2}{*}{\centering }} & \multicolumn{1}{c}{\multirow{2}{*}{\centering }} & \multicolumn{1}{c}{\multirow{2}{*}{\centering }} & \multicolumn{1}{c}{\multirow{2}{4cm}{\centering Lexicographic obj., ECMP routing}} \\
\\ 
\hline 
\multicolumn{1}{c}{\multirow{2}{*}{\centering \cite{coiro14}}} & \multicolumn{1}{c}{\multirow{2}{*}{\centering x}} & \multicolumn{1}{c}{\multirow{2}{*}{\centering }} & \multicolumn{1}{c}{\multirow{2}{*}{\centering }} & \multicolumn{1}{c}{\multirow{2}{*}{\centering }} & \multicolumn{1}{c}{\multirow{2}{*}{\centering }} & \multicolumn{1}{c}{\multirow{2}{*}{\centering }} & \multicolumn{1}{c}{\multirow{2}{*}{\centering }} & \multicolumn{1}{c}{\multirow{2}{*}{\centering }} & \multicolumn{1}{c}{\multirow{2}{4cm}{\centering Node expansion – Multiple sleeping states}} \\
\\
\hline 
\cite{polverini15} & x &  &  &  &  &  &  &  & New power state  \\
\hline 
\cite{phillips11} &  &  &  & x &  &  &  & x &  \\ 
\hline 
\cite{moulierac15} &  &  &  & x &  &  &  &  & ECMP and robustness  \\ 
\hline 
\cite{lee12b} &  &  &  &  &  &  &  &   & Fully splittable \\ 
\hline 
\cite{shen12} &  &  &  &  &  &  &  &   & free splitting ratio \\
\hline
\cite{francois13a,francois14a} &  &  &  &  &  &  &  &  &  \\ 
\hline
\multicolumn{1}{c}{\multirow{2}{*}{\centering \cite{bianzino12b,bianzino12c}}} & \multicolumn{1}{c}{\multirow{2}{*}{\centering }} & \multicolumn{1}{c}{\multirow{2}{*}{\centering }} & \multicolumn{1}{c}{\multirow{2}{*}{\centering }} & \multicolumn{1}{c}{\multirow{2}{*}{\centering }} & \multicolumn{1}{c}{\multirow{2}{*}{\centering }} & \multicolumn{1}{c}{\multirow{2}{*}{\centering x}} & \multicolumn{1}{c}{\multirow{2}{*}{\centering }} & \multicolumn{1}{c}{\multirow{2}{*}{\centering x}} & \multicolumn{1}{c}{\multirow{2}{4cm}{\centering Traffic unaware because only link loads considered}} \\
\\ 
\hline 
\cite{cianfrani10,cianfrani11} &  &  &  &  &  & x &  & x & OSPF is modified \\ 
\hline 
\cite{cianfrani12} &  &  &  &  &  &  &  & x & OSPF is modified \\ 
\hline 
\cite{lee13} &  &  &  &  &  &  & x & x & Single node failures \\ 
\hline 
\label{tab:problem_features_sp}
\end{tabular}

\end{scriptsize}
\end{table}

\end{landscape}

The second axis of classification are the resolution approaches.
Tables \ref {tab:approach_type_fb} and \ref{tab:approach_type_sp} show,
for the flow-based and shortest path based routing, respectively, which of
the articles deal with centralized, distributed, exact, heuristic, on-line or off-line algorithms. 
We can see that for both types of routing, the majority of the proposals are centralized and on-line.
Heuristic approaches are used in an overwhelming manner and they are the totality of
the work that deal with shortest path routing.

\renewcommand\arraystretch{1.1}
\begin{table}[t!bp]
\centering
\caption{Literature classification according to resolution approaches: flow based routing}
\begin{tabular}{ccccccc}

{\bf Features} & Centralized & Distributed & Exact & Heuristic & Online & Offline \\ 
\hline 
{\bf Work} & & & & & & \\
\cite{addis14} & x &  & x (MILP) & x  & x & x \\ 
\hline 
\cite{addis12b} & x &  &  & x  &  & x \\ 
\hline 
\cite{addis12,addis12a} & x &  & x (MILP) & x  &  & x \\ 
\hline 
\cite{addis13} & x &  & x (MILP) & x  &  & x \\ 
\hline 
\cite{addis14a} & x &  & x (MILP) & x  &  & x \\ 
\hline 
\cite{amaldi13c} & x &  & x (MILP) & x  &  & x \\ 
\hline 
\cite{chabarek08} & x &  & x (MILP) &  &  & x \\ 
\hline 
\cite{zhang10} & x &  &  & x  & x &  \\ 
\hline 
\cite{sanso09} & x &  & x (MILP) & & & x \\
\hline
\cite{chiaraviglio12a} & x &  & x (MILP) & x  & x &  \\ 
\hline 
\cite{chiaraviglio08,chiaraviglio09,chiaraviglio09a,lee12a,cheung14} & x &  &  & x  & x &  \\ 
\hline 
\cite{garroppo13a,garroppo13} & x &  & x (MILP) & x &  & x \\ 
\hline 
\cite{garroppo11,garroppo12,giroire10,fisher10,lin12,lin14,lin13,mumey12,wu13,galan-jimenez13a,galan-jimenez13b} & x &  &  & x  & x &  \\ 
\hline 
\cite{charalambides13} &  & x &  & x  & x &  \\ 
\hline 
\cite{kist11} & x &  & x (MILP) &  &  & x \\ 
\hline 
\cite{cianfrani12a,coiro13a,cianfrani13} & x &  &  & x  & x &  \\ 
\hline 
\cite{vasic10} &  & x &  & x  & x &  \\ 
\hline 
\cite{vasic11} &  & x &  & x  & x & x \\ 
\hline 
\cite{avallone12,seoane11} & x &  &  & x  & x &  \\ 
\hline 
\cite{kim11,kim12} &  & x &  & x  & x &  \\ 
\hline 
\cite{coiro13} &  & x &  & x  & x &  \\ 
\hline 
\cite{takeshita12} & x &  &  & x  & x &  \\ 
\hline 
\cite{hou14,yang15} &  & x &  & x  & x &  \\ 
\hline 
\cite{francois12,francois13} & x &  &  & x  &  & x \\ 
\hline 
\cite{francois14} &  & x &  & x  & x (TE) & x (PATH) \\
\hline 
\cite{bianzino12a} & x &  &   &x  &x  &  \\ 
\hline 
\cite{zhao13} & x &  &  & x  &  & x \\ 
\hline 
\cite{giroire12,koster13,coudert13} & x &  &  & x  &  & x \\ 
\hline 
\cite{niewiadomska-szynkiewicz13,niewiadomska14} & x &  & x (MILP) & x  & x &  \\ 
\hline 
\cite{luo13} & x &  & x (MILP) &  &  & x \\ 
\hline 
\cite{aldraho12} & x &  & x (MILP) &  &  & x \\ 
\hline
\cite{lin15} & x & & & x & x & x \\
\hline 

\end{tabular}
\label{tab:approach_type_fb}
\end{table}

\begin{table}[t!bp]
\centering
\caption{Literature classification according to resolution approaches: shortest path routing}
\begin{tabular}{ccccccc}
{\bf Features} & Centralized & Distributed & Exact & Heuristic & Online & Offline \\ 
\hline
{\bf Work} & & & & & & \\ 
\cite{amaldi13,amaldi11,amaldi11a} & x &  &  & x &  & x \\ 
\hline 
\cite{capone13} & x &  &  & x & x &  \\ 
\hline 
\cite{coiro14} & x &  &  & x  & x &  \\ 
\hline 
\cite{polverini15} & x &  &  & x  & x &  \\ 
\hline 
\cite{phillips11} & x &  &  & x &  & x \\ 
\hline 
\cite{moulierac15} & x &  &  & x &  & x \\ 
\hline 
\cite{lee12b} & x &  &  & x & x & x \\ 
\hline 
\cite{shen12} & x &  &  & x & x & x \\ 
\hline 
\cite{francois13a,francois14a} & x &  &  & x & x & x \\ 
\hline 
\cite{bianzino12b,bianzino12c} &  & x &  & x & x &  \\ 
\hline 
\cite{cianfrani10,cianfrani11} & x &  &  & x & x & x \\ 
\hline 
\cite{cianfrani12} & x &  &  & x & x & x \\ 
\hline 
\cite{lee13} & x &  &  & x &  & x \\ 
\hline
\end{tabular}
\label{tab:approach_type_sp}
\end{table}

\subsection{Detailed description}
\subsubsection{Flow-based Routing}
Some of the most popular studies on EANM, \cite{chiaraviglio08, chiaraviglio09,chiaraviglio09a,chiaraviglio12a}, explicitly address the variant of the IP-BEANM problem (\ref{eq:ob.base}-\ref{eq:domain_real.base.nodes}) adapted to model sleeping capabilities (see Section \ref{sec:sleeping}). We recall that  (i) routing is flow based and fully splittable, that is, flow variables $f_{ij}^{d}$ are real and nonnegative, (ii) the objective function minimizes the fixed energy cost paid to keep a node (router) or a link (network interface) in the active state and no proportional consumption component is considered, i.e., objective function (\ref{eq:ob.base}) is replaced by (\ref{eq:ob.sleep}), (iii)  link capacity constraints (\ref{eq:capacity.base}) are adjusted as in (\ref{eq:capacity.sleep}), (iv) each link is assumed to be made by a single network interface or line card, i.e., $n_{ij} = 1,\, \forall (i,j) \in A$. The MILP formulation presented in \cite{chiaraviglio08, chiaraviglio09,chiaraviglio09a} is solved only in \cite{chiaraviglio12a}, where the authors propose some little refinements to reduce the MILP complexity, as switching to a \emph{per-source routing scheme}. Note that the proposed MILP presents an alternative formulation for the node activation constraints (\ref{eq:coherence1.sleep}-\ref{eq:coherence2.sleep}), which is based on the use of a big-M parameters $M$:

\begin{equation}
\label{eq:coherence.bigM}\sum_{(i,j)\in A}w_{ij} + \sum_{(j,i) \in A} w_{ji} \leq M y_{i},\qquad \forall i \in V. 
\end{equation}

Note that this formulation is dominated by  (\ref{eq:coherence1.sleep}-\ref{eq:coherence2.sleep}), and therefore is less effective. Furthermore the choice of parameter $M$ can greatly affect the efficiency of the formulation.

The main contribution of these papers is a standard greedy algorithm which, after sorting nodes and links according to a specific criterion, tries to put to sleep each element of the ordered list. The inputs of the procedure are the traffic matrix, the network topology with link capacities, and a set of link weights (considered only by the algorithm, not by the routing protocol) used to determine the network routing according to a shortest path scheme. %\nota{qui parliamo di shortes path, ma non e' quello delle shortest path routing, giusto? Forse va spiegato in due righe a cosa serve. Mi pare sia spiegato dopo} 
A network element is put to sleep only if the weight set of the active elements determines a feasible routing which satisfies maximum utilization constraints (\ref{eq:capacity.sleep}). In \cite{chiaraviglio08,chiaraviglio09,chiaraviglio09a,chiaraviglio12a} different sorting policies are progressively proposed and tested.

The presented algorithms are intended to be run by a centralized management platform in control of the whole network and able to estimate the current traffic matrix. Furthermore, due to the very-low time requirements, they are meant to be performed periodically, in real-time, about every thirty minutes. The authors do not explicitly investigate the issues of stability and re-configuration frequency. 

Another early work to deal in a centralized manner with the IP-BEANM problem is \cite{sanso09}, where three MILP formulations with fully splittable routing are proposed, solved and evaluated in terms of performance and network survivability. Both flow proportional and fixed power consumption components are considered, and a bi-objective function is designed to find the trade-off between energy-savings and network performances.

The same problem addressed in \cite{chiaraviglio09} is considered in \cite{lee12a}, where the authors propose a heuristic centralized approach to improve upon \cite{chiaraviglio09}. The method first exploits a state-of-the-art steiner-tree algorithm to compute the minimal set of nodes to be switched on to guarantee the network connectivity (all the links connecting activated nodes are kept on at this stage). Sleeping nodes are then progressively activated, together with their links, until the maximum utilization constraints (\ref{eq:capacity.sleep}) are satisfied. Finally, a scheme very similar to that used in \cite{chiaraviglio08,chiaraviglio09,chiaraviglio09a,chiaraviglio12a} allows to put to sleep the largest possible set of links. The feasibility test that verifies link utilization constraint violation is done by computing K-shortest paths for each traffic demand, and by successively running a state-of-the-art bin packing algorithm to choose the best routing paths. The proposed procedure is meant to be run online every five minutes to dynamically adapt the network configuration to the incoming traffic.

A variant of the greedy algorithm of \cite{chiaraviglio09} is presented in \cite{cheung14}, where multiple shortest paths are tested for each traffic demand to verify whether a node deactivation would cause congestion.

A similar centralized online approach is presented in \cite{giroire10} where, however, only links (line cards) can be put to sleep. In addition to an extensive analysis on the problem complexity, the authors propose another greedy heuristic very similar to those just discussed. In this case, the feasibility test is done by sequentially routing each demand on the shortest path defined by the ratio between the residual and the full capacity. A centralized online strategy based on a greedy algorithm for the IP-BEANM problem with sleep-capable links is illustrated in \cite{bianzino12a}: the link sorting criteria is the link importance within the network, which is represented by its \emph{Shapley value}.

Flow-proportional convex energy costs related to the so-called \emph{route-processor} are included in  \cite{garroppo11,garroppo12}. A greedy approach is  again chosen to solve the problem. In \cite{garroppo11} both nodes and links can be put to sleep. The heuristic scheme is the same of \cite{chiaraviglio09}, except for the feasibility test, which is performed by solving an LP formulation to compute a routing configuration which does not violate the link capacity. In \cite{garroppo12}, only links can be put to sleep, and multiple line cards are considered on each link, i.e., $n_{ij}\,\geq 1$ (bundled links).
The same greedy scheme of \cite{garroppo11} is adapted to cope with the new problem. At each iteration a single line card is considered for the switching off, and the Fast Greedy Heuristic presented in \cite{fisher10} is solved to avoid maximum utilization constraint violation. These approaches are centralized and could be applied online as they have small computational time.

The formulations presented and heuristically tackled in \cite{garroppo11,garroppo12} are later solved in \cite{garroppo13,garroppo13a}. The convex non-linear energy terms are linearized trough a piece-wise linear approximation (see, e.g., \cite{fortz02}). The resolution approach requires a substantial amount of time and is only suitable for offline optimization.
In \cite{takeshita12} the authors solve a IP-BEANM problem with sleep capable links, nodes with proportional power profile, single path routing and additional QoS constraints, i.e., maximum utilization constraints, maximum hop number, and what the authors call disjoint multi-route divergence. The procedure is based on brute force generation and testing the largest possible number of routing configurations.

A different perspective to address the IP-BEANM problem with both fixed and proportional energy costs is assumed in \cite{vasic10} and \cite{vasic11}. In \cite{vasic10} an online distributed procedure according to which each edge router (those generating traffic) is periodically responsible for adjusting the amount of traffic sent through the pre-defined transmitting paths: the aim is to minimize the consumption of the network links. The distributed procedure is equivalent to a classic local-search algorithms. Different shifting moves are included according to the shape of the device power profiles, which can be ON-OFF prevalent or step-wise proportional ({ALR}-likewise). Since pre-defined paths are used, the corresponding optimization problem should consider a \emph{per-path} routing scheme (see Section \ref{sec:perpath}). A second distributed approach of the same authors to solve the same optimization problem is presented in \cite{vasic11}. In this case the pre-defined set of paths is optimized in advance with the aim of identifying those \emph{critical} paths that should always be kept  on to guarantee the connectivity while using the minimal number of links. A new on-line distributed architecture called REsPoNse is then responsible to adjust the portion of traffic sent on each pre-computed path. The set of pre-computed paths is split among critical always-on paths and on-demand paths that can be exploited in case of congestion or device failures. 

In \cite{avallone12} an admission control scheme to put to sleep both links and nodes is proposed. Once a flow demand (traffic demand) is received from the network controller, the less consuming feasible single path is computed and assigned. From the optimization methodology perspective, these framework implements a greedy scheme which sequentially route each traffic request $d \in D$. Note that in case of saturation some traffic demands may not be accepted into the system, which is equivalent to accepting an unfeasible solution for the corresponding optimization problem. A very similar greedy scheme is adopted in \cite{seoane11} to solve the basic EANM problem: traffic demands are sorted and routed one by one on the best path, in terms of incremental energy consumption, chosen from a set of precomputed K-shortest paths.

In \cite{coiro13} the same optimization problem is addressed by means of a distributed architecture called DAISES for IP/MPLS. The framework dynamically adjusts the single path (the Label Switched Path) of each traffic demand by computing the shortest-paths determined by a set of link weights which are both energy consumption and congestion related. Energy/congestion related weights are considered in \cite{hou14} too, where, w.r.t. \cite{coiro13} the optimization problem assume routing to be fully splittable. The weights are exploited by  the proposed hop-by-hop routing protocol to practically implement, in a distributed way, a variant of the well-known flow-deviation method \cite{fratta73}. Hop-by-hop routing mechanisms are also discussed in \cite{yang15}, where the goal is to dynamically minimize the link consumption in a MLR scenario by considering energy related link weights to choose the less consuming route toward the destination.

 Another distributed routing mechanism is proposed in \cite{kim11} and \cite{kim12} to dynamically minimize link and node consumption while exploiting the concept of \emph{traffic centrality}, which, given a node $i \in V$, is the ratio between the traffic carried by the most loaded outgoing link and the total outgoing traffic.  The authors claim that maximizing this amount for each node is equivalent to minimizing power consumption, however that holds only under very specific conditions. Each node chooses with a certain probability the interface used to forward each flow. The probabilities change according to delay measures taken by ant probes transmitted through the network. In this case the routing is fully splittable since this probability-based forwarding exploits all the outgoing interfaces whose choice probability is greater than zero. Furthermore, the method is traffic unaware because no traffic matrix is required.

The IP-BEANM problem adapted for the bundled links case with fully splittable routing (see Section \ref{sec:bundled}) has been largely addressed in the literature. If no specified otherwise, the next frameworks considering with bundled links are all meant to be employed in a centralized online fashion. In \cite{fisher10}, which was the first work to appear, the authors present three different heuristic algorithms, i.e., Fast Greedy Heuristic (FGH), Exhaustive Greedy Heuristic (EGH) and Bi-level Greedy Heuristic (BGH), to quickly determine the sub-optimal set of line cards to be put to sleep. To this purpose, a specific LP formulation to maximize the network spare capacity is solved at each iteration to verify whether one of the active cables of the bundle with more spare capacity can be switched off without saturating the network. EGH and BGH involve some variants aimed at avoiding local optima. 

The same problem with single-path routing (see Section \ref{sec:sp}) is addressed in \cite{lin13}, where two proposed procedures are based on the computation of the K-shortest-paths of each source-destination pair. These paths are exploited to determine, at each iteration, if there exist a routing configuration which satisfy maximum utilization constraints every time a cable is put to sleep. In other two articles \cite{lin12} and \cite{lin14}, the same authors consider another problem variant, according to which nodes can be put to sleep, routing is fully splittable and a pre-computed set of paths is given as input to the model (\emph{per-path} routing). The formulations presented in \cite{lin12} and \cite{lin14} contain several flaws and fail to correctly identify the real problem addressed. The input path set is composed of the K-shortest paths that satisfy each demand $d \in D$. The same scheme already used in many approaches already mentioned is used to put to sleep the elements. Every time a cable or a node is tested for sleeping, a heuristic procedure is called to find an alternative feasible routing on the residual network. Note that in \cite{lin14} the pre-computed paths cannot be longer than a certain threshold, but since paths are provided in input, that is not translated into an additional group of constraints for the corresponding optimization problem.

Another variant with sleeping nodes, sleeping cables and classic \emph{per-flow} fully splittable routing is considered in \cite{mumey12}, where, again, shortest-path-based and tree-based algorithms are combined to maximize the number of sleeping nodes. Though we consider only IP problems, let us cite \cite{wu13}, where a pseudo-multi-layer IP over {WDM} variant of the classic bundled links problem is addressed. We refer to this work as \emph{pseudo-multi-layer} because the WDM elements is introduced by simply naming the cables of each bundle as logical links of the optical layer: there is no fundamental difference with respect to \cite{mumey12}. A {MILP} formulation and a {LR} heuristic are presented and tested.

Line cards formed by multiple ports (equivalent to bundled links) are considered in \cite{zhang10}, where a complete formulation for the IP-BEANM problem with fixed energy cost on line cards and network ports, fully splittable routing and constraints on maximum utilization is reported. To rapidly find a sub-optimal solution the authors propose to modify the base MILP and switch to a \emph{per-path} routing based on K precomputed paths. These paths are selected to respect constraints on the maximum number of hop or on the maximum propagation delay. The methodology is meant to be applied online every 15 minutes in a centralized manner.

A very different approach with respect to the literature already mentioned is adopted in \cite{charalambides13}, where the IP-BEANM problem with bundled links is handled in a distributed manner: each node is allowed to decide the splitting ratios within a pre-defined set of paths. Note that, although this work is based on multi-topology routing, with each topology defined by a specific set of link weights and shortest path routing over each single topology, the high level abstraction of the problem does not fall in the class of shortest path routing, but in that of flow-based \emph{per-path} one: for each demand a pre-defined set of paths determined by the multiple virtual topologies is given as input, and the framework decides which path should be chosen.

Another distributed online approach for the IP-BEANM problem with bundled links is presented in \cite{francois14}. The authors propose to use the backup path already set up by the MPLS protocol to aggregate the traffic on a smaller number of network line cards. Each router can independently choose if using the primary or the backup path to forward the flowing demands. From the optimization model perspective, this framework heuristically solves a \emph{per-path} version of the IP-BEANM problem, with a couple of available paths for each demand, i.e., the primary and the backup one. Some tabu parameters are used to forbid shifting the flow on a link already impacted by other shifting moves.

A new online centralized management architecture for IP/MPLS networks relying on novel optimization algorithms to optimize device states (routers, ports and links) by considering single-path routing is described in \cite{niewiadomska-szynkiewicz13} and \cite{niewiadomska14}. Multiple power states are defined for each link, each one characterized by specific capacity and consumption values. Both a standard \emph{per-arc} and a \emph{per-path} formulation based on pre-computed paths are given. A heuristic built on the linear relaxation of an incremental model and progressive variable fixing are proposed.

Differently from a bundled link configuration, when ALR is considered, the network manager has to decide the operational rate of each link: in \cite{galan-jimenez13a,galan-jimenez13b}, the authors evaluate how the number of operational rates available (energy-levels in the papers) impacts on the potential energy savings. To solve the BEAN problem adapted as shown in Section \ref{sec:alr} to account for ALR, the authors propose two novel heuristic procedures based, respectively, on a genetic algorithm and on particle swarm optimization. 

A novel perspectives in flow-based EANM problems is introduced by \cite{kist11,cianfrani12a,coiro13a} and \cite{polverini15}, which consider and model different power states of routers and line cards in addition to the classic ON-OFF. In~\cite{kist11} is presented the first modeling of special stand-by states, according to which only some routing/forwarding functions are available or not. Besides sleeping and active states, the authors introduce the following operational states for network routers (nodes): (i) \emph{bridged-all}, (ii) \emph{bridged-local}, (iii) \emph{default-gateway} and (iv) \emph{bridged-many}. An exact {MILP} formulation is presented and evaluated. This approach is suitable for a centralized offline implementation.

The same problem is addressed in a heuristic fashion in \cite{cianfrani12a}, where a procedure based on the Floyd-Warshall algorithm is proposed. A power state for routers is introduced in \cite{cianfrani13}, while the new stand-by states are applied to line-cards in \cite{coiro13a}. To solve all the problems the authors present a novel ad-hoc heuristic. Either \cite{cianfrani12a}, \cite{cianfrani13} and \cite{coiro13a} could be applied online in a centralized way.

The first work dealing with a very special multi-period optimization problem is \cite{francois13} (preliminary version in \cite{francois12}), who propose a centralized off-line approach aiming at computing a low consumption configuration to be applied during the low traffic periods, and a more consuming one required during the peak phases. Multi-periodicity is here addressed in a different way with respect to what is shown in Section \ref{sec:mp}, since one of the variables of the optimization problem is the duration of the applicability period for the low consumption configuration. The proposed heuristic first follows a greedy mechanism like that of \cite{chiaraviglio09} to put to sleep the links for the less loaded traffic matrix. Then the links are progressively re-added into the active topology to guarantee capacity constraints for the more loaded traffic matrices and, finally, the algorithm chooses the low traffic configuration according to the product between the energy consumption and the applicability window duration. The algorithm is adjusted to account also for single link failure protection: during the greedy procedure it is necessary to check whether, after the elimination of a link, the current link weight setting is still able to satisfy capacity constraints in case of failure of any other active links.
To the best of our knowledge, the only other work to explicitly address a multi-period IP-BEANM problem with flow-based routing are \cite{addis12b} and \cite{addis14}. In \cite{addis14}, two MILP exact formulations are presented, the first  assumes a fixed routing configuration along the whole set of time periods, and the second allowing the network administrator to modify the routing path at the beginning of each new time interval. Multiple line cards are considered on each link (bundled links), maximum utilization constraints are imposed, energy savings are achieved by putting to sleep line cards and router chassis and by limiting the reactivation of the latter (reactivation causes a consumption spike). Furthermore, limitations on the maximum number of times that a single line card can be switched on within the whole set of periods are enforced to preserve the line card lifetime (see Section \ref{sec:mp}). In \cite{addis12b} a novel heuristic to quickly find a sub-optimal solution for the two previous formulations is proposed. The algorithm is based on a GRASP methodology which sequentially solves a modified version of the basic formulation which considers a single traffic demand. Another heuristic is presented in \cite{addis14}, where the configuration of each single time period is obtained by solving a single-period version of the reference MILP. While the exact MILP and the GRASP-based algorithm are intended for offline centralized approaches only, the single period heuristic can be applied in an online fashion, by solving the single-period problem at the beginning of each time interval or every time the network becomes too saturated or too empty.

While there is a substantial literature on the integration, at the optical layer, of protection techniques with energy-aware network management practices (see the related work section in \cite{addis14a}), only a few studies have dealt with this problem at the IP layer \cite{aldraho12,luo13,addis12,addis12a} and \cite{addis14a}. In \cite{aldraho12} two MILP formulations to provide protection to single link failures while minimizing the node and link consumption are presented. The first implements dedicated protection (see Section \ref{sec:pr}) while the second considers a less usual protection scheme, i.e., link protection, according to which a backup path is defined between each couple of nodes connected by a link. In both case routing is single path, both fixed and flow proportional components are considered  present different MILP formulations to reduce network consumption while guaranteeing link protection or demand protection. Shared protection in a multi-rate scenario characterized by six different rate/energy states for each device (either routers, either links) is discussed in \cite{luo13}. Three MILP formulations considering protection are presented, the first two based on dedicated protection and the last one on shared protection. One of the formulation for dedicated protection assumes that backup paths have no impact on consumption. Note that the notation used in the paper is not clear and several flaws are present. The formulation proposed in \cite{aldraho12} and\cite{luo13} can be applied offline in a centralized manner.

In the preliminary work presented in \cite{addis12} and \cite{addis12a}, the multi-period problem formulations discussed in \cite{addis12b} and \cite{addis14} are integrated to support, respectively, dedicated protection and shared protection. Both papers propose a MILP formulation and adapt the single-period heuristic previously mentioned for the unprotected problem to the novel protected ones. The authors compare a \textit{classic} approach where all network elements which carrying at least a primary or a backup path are powered-on, and a \textit{smart} version which allows the switching-off of line cards crossed by backup paths exclusively. Furthermore, to further reduce the network consumption, a second utilization threshold (larger than the main one) to be respected during failure periods only is introduced. 
The problem addressed \cite{addis12b} and \cite{addis14} is further extended in \cite{addis13} to account for robustness to traffic variations. The base formulation is modified to integrate the cardinality constrained method first proposed in \cite{bertsimas11} to manage traffic demand which are uncertain within a close symmetric interval.
The work presented in \cite{addis12,addis12a} and \cite{addis13} are finally integrated into a comprehensive formulation \cite{addis14a} which is exploited to conduct an overall trade-off evaluation between energy-efficiency and network survivability. All the above formulations are suitable for a centralized offline approach, while the single-time period heuristic adapted to each specific problem variant can be employed in an online manner.

A different strategy to provide network resilience in a centralized fashion is proposed in \cite{lin15}, where the authors aim at keeping the failure probability for each traffic demand under a given threshold. The failure probabilities for a demand is related to the failure probability of each single link crossed by the routing path, and by the number of cables which constitute each link itself. Two heuristic algorithms based on the computation of K-shortest paths for each demand (similar to \cite{lin14}) to put to sleep link cables are presented and tested. Both algorithms are suitable for on-line and off-line implementation.

Finally, let us mention some work that adopt less conventional methodologies. The so-called \emph{Nash-bargaining} model is considered in \cite{zhao13} to solve in a centralized and offline fashion an only-link IP-BEANM problem adjusted to account for a bi-objective function, where both energy consumption and congestion costs are minimized. Only flow proportional consumption components are considered. The authors define and solve a convex programming formulation. In \cite{giroire12,coudert13} and \cite{koster13}, the authors address a very particular IP-BEANM problem to model the redundancy elimination mechanism, according to which, a subset of routers equipped with enhanced and power hungry capabilities can store the content of the transmitted packets: every time the same content has to be retransmitted toward another "capable" router, the routers does not send a full packet, but a simple hash which allows the receiver router to identify the content in its memory. This mechanism produces a bandwidth reduction, which can be exploited to save energy. In \cite{giroire12}, a MILP formulation and a heuristic algorithm to put to sleep nodes and links while assuming fully splittable routing and deciding whether to activate the redundancy elimination function are presented. New cutset inequalities for the same problem are presented in \cite{koster13}, while in \cite{coudert13} the robust version of the problem is formulated. As in \cite{addis13} and \cite{addis14a}, the robust MILP exploits the cardinality-constrained approach formulated in \cite{bertsimas11} to balance the solution robustness. All the proposed approaches are suitable for offline centralized management mechanisms.
In \cite{amaldi13c}, a novel bi-level centralized optimization approach for EANM with elastic traffic demand (see Section \ref{sec:elastic}) is presented. The authors present an exact MILP formulation with \emph{per-flow} single path routing and a heuristic MILP based on a precomputed set of paths. Energy savings are achieved by putting to sleep network links and routers. The computation of the correct transmission rate to be assigned to each elastic flow is performed by introducing a group of constraints which model a max-min-fair allocation. Due to the high computing times the methodology can be applied offline only.
To conclude the part dedicate to flow based routing, let us cite the very popular work of Chabarek et al. \cite{chabarek08}, which is usually used as reference to prove the prevalence of the fixed consumption components in traditional network device. In this paper a MILP for joint energy-aware network design and routing with flow based splittable routing is proposed and solved.

\subsubsection{Shortest Path Routing}
The additional constraints and variables required to correctly determine shortest paths, i.e., (\ref{eq:commonflow.sp}-\ref{eq:domain_real.sp}), make EANM problems with shortest path routing much more complex than flow-based ones. Thus, even though from the practical perspective adjusting the link weights of the shortest path routing protocol seems a very straightforward way to perform traffic engineering, the complexity issue has made this optimization problem less tractable and less attractive for network scientists.
In this field the literature is less rich, but still offers some very interesting proposals.

In \cite{amaldi11,amaldi11a} and \cite{amaldi13}, the authors address an EANM problem with shortest path routing with equal cost multi-path, where both links and routers can be put to sleep and $n_{ij}\,=\,1$ for each link $(i,j) \in A$. The heuristic solution first proposed in \cite{amaldi11}, has been refined first in \cite{amaldi11a} and finally in \cite{amaldi13}, where a comprehensive centralized offline algorithm to obtain an energy-aware weight setting is described and tested. A peculiarity of this work is that once energy consumption is minimized, the proposed procedure then focuses on minimizing a measure of network congestion too: more technically, it is equivalent to say that the algorithm lexicographically optimizes, in the following order, energy consumption and network congestion. The framework exploits (i) a greedy procedure similar to that used in \cite{chiaraviglio09} to quickly put to sleep the more trivial network elements, solves a relaxed version of the corresponding EANM problem where routing is kept flow-based to deactivate the more crucial ones, and (iii) run the IGP-WO algorithm proposed in \cite{fortz02} to optimize the link weights of the reduced topology.
The same algorithm is exploited in \cite{capone13} to build a joint offline-online optimization platform to solve a multi-period version of the problem where the goal is to optimize the configuration along an entire day: multiple sets of energy-aware weights are computed during a planning phase by the offline algorithm, and then dynamically applied in the network by the smart network controller according to real-time traffic measurements.

Multi-period offline energy-aware link weight optimization with sleep-capable links is addressed in\cite{phillips11}, where similarly to \cite{capone13} the authors propose to determine a finite set of energy-aware configurations and their applicability window for {OSPF}-operated {IP}. A first genetic algorithm is run to compute a sub-optimal network configuration for a certain level of traffic, while a second genetic algorithm is employed to determine when and fr how long each configuration will have to be applied to minimize the whole consumption and avoid too frequent re-configurations. Differently from \cite{capone13}, only the weights of sleeping elements are adjusted (with very high values which force the exclusion from any shortest path tree), and the addressed problem is explicitly  exploited to find the sub-optimal configurations that satisfy an estimated traffic matrix representing specific levels of traffic. A single configuration determines the states of link and routers. Note that in this work, the link weights of active elements are kept unchanged. Furthermore, configuration switching are optimized and planned in advance.

A multi-period scenario similar to that addressed in \cite{addis14} but based on shortest path routing is addressed in \cite{moulierac15}. The authors present a novel greedy approach to put to sleep network links ($n_{ij} = 1$) while minimizing the number o weight changes across the different time periods of the day. As in \cite{phillips11}, only the link weights of sleeping elements are modified (with very high values). Furthermore, robustness to traffic variations is included into the framework as previously done in \cite{addis13}. The configuration of each time period is computed separately, starting from the that of the less to that of the more loaded. The method is suitable for off-line centralized planning.

Another genetic algorithm to solve in a centralized offline manner a joint optimization problem to minimize both network consumption and maximum link utilization is presented in \cite{francois13a} and \cite{francois14a}. Although the authors claim to solve a classic EANM problem with multicommodity flow (flow-based routing), the proposed procedure is based on the optimization of the link weights of an Interior Gateway Protocol: for this reason we place this work in the shortest-path routing section.

The same problem of \cite{amaldi13} where only links can be put to sleep is solved by the set of optimization approaches described in \cite{lee12b}. Starting from a \emph{per-path} formulation of the problem, the authors build a Lagrangian relaxation which is successively used as pillar of three heuristic algorithms: (i) the {LR} algorithm which progressively updates the dual multipliers of capacity constraints and finally converts them into link weights, (ii) the Harmonic Series (HS) algorithm based on a local search procedure to update and improve an input set of link weights and (iii) a combined LRHS heuristic where the weights computed by the LR algorithm are given as input to the HS. The authors affirm that the proposed algorithms, which were originally  meant for off-line centralized optimization, are fast enough to be integrated in an on-line optimization framework too.
The LR of the main formulation is also used in \cite{lee13}  by some of the same authors of \cite{lee12b} to solve a protected  version of the problem where single node failure resilience is provided through optimized link weights.	

The heuristic link weight optimization method proposed in \cite{shen12} to jointly maximize the number of unused links and minimize the maximum link utilization assumes that network operators can freely configure the splitting ratios among multiple shortest-paths. 

Another work considering shortest path routing but without the possibility of optimizing the link weights has been very recently proposed in \cite{coiro14}. The authors consider the table-look operation first discussed in \cite{coiro13a} with flow-based routing, and present a new ad-hoc genetic algorithm to solve the energy minimization problem with shortest path routing. The same proposal is successively refined in \cite{polverini15}, where the authors define a new router power state called $F^3$, define and solve the corresponding MILP formulation to decide which router should be put in $F^3$ state to minimize energy consumption, and present a heuristic centralized algorithm to deal with larger networks.

A distributed online strategy follows the frameworks proposed in \cite{bianzino12c} and \cite{bianzino12b}. Link load information disseminated through the networks is used, according to a given policy, by the whole set of routers \cite{bianzino12c}, or by each single router in an independent manner \cite{bianzino12b}, to periodically put to sleep a candidate link. If this operation causes other links to become saturated, it is immediately reversed.

Finally, let us mention the Energy Aware Routing (EAR) algorithm first presented in \cite{cianfrani10} and successively enhanced in \cite{cianfrani11,cianfrani12}, which address the EANM problem by presenting a heuristic based on a restated version of the {OSPF} protocol. 
EAR discriminate among Importer Routers (IR), that are exempted from computing their own shortest path trees (SPTs) and Exporter Routers (ER) which provide the SPTs to the neighboring IRs. This modification of the protocol allows to use a smaller number of active SPTs, which would naturally increase the number of links not belonging to any shortest path, and thus eligible for sleeping. In \cite{cianfrani10} and \cite{cianfrani11} the framework is traffic unaware and is meant to be applied during night hours only, while in \cite{cianfrani12} traffic matrix knowledge is introduced to guarantee the quality of service. The optimization problem which is directly solved the choice of IRs and ERS which maximizes the energy savings. Practically speaking, a heuristic is obtained by modifying the terms of the problem.

\subsubsection{Other Routing Schemes}
For the sake of completeness, here we mention a few other interesting works (not reported in the summary tables) which do no deal with neither flow-based nor shortest-path routing.
Some traffic-unaware heuristic algorithms to choose the sleeping elements are described in \cite{cuomo11,cuomo11a,cuomo12}. The authors do not refer to any specific routing scheme and base their approach on topological features exclusively, e.g., the number of shortest paths crossing each link. The idea is to remove from the network the elements which seem to be less critical from connectivity perspectives. Traffic unawareness makes this centralized offline approach suitable for very low traffic conditions only.
In \cite{capone12} the authors consider a tree-based routing scheme adopted in network operated with Carrier Grade Ethernet. Two different bi-objective formulations to balance energy consumption and network performance are proposed. Both nodes and links are put to sleep. The trees used to define the possible paths are pre-computed and provided as input to the models.

\section{Conclusions}

An important problem inside the green networking literature is the EANM, the energy-aware network management problem that deals with
reducing network consumption by managing the different resources of the network. 
This paper has given a tutorial view of what are the major issues that influence the different EANM proposals.
Next, we have explored in-depth the significance of those features in rigorous mathematical programming approaches and compared them
with other classical modeling features in the operations research literature. 
The next step was to classify the large body of literature according to the problems features or the resolution approaches and, finally,
we provided a detailed explanation of each of the proposals that could be found up to date. 
We have found commonality in the different approaches and pinpointed those that are particularly rare or present specific features.
We hope that this tutorial-survey that we have called a ``journey through optimization glasses" can shed some light
into the complex world of energy-aware network management.

\vspace{1cm}

\end{document}